\documentclass[american,aps,pra,reprint, superscriptaddress, twocolumn]{revtex4-1}
\usepackage{amsmath,amscd,amsthm,amssymb}
\usepackage{mathrsfs}
\usepackage{graphicx}
\usepackage{epsf,epstopdf}
\usepackage[all]{xy}
\usepackage[unicode=true,pdfusetitle, bookmarks=true,bookmarksnumbered=false,bookmarksopen=false, breaklinks=false,pdfborder={0 0 0},backref=false,colorlinks=false] {hyperref}
\hypersetup{colorlinks=true,linkcolor=myurlcolor,citecolor=myurlcolor,urlcolor=myurlcolor}
\usepackage{graphics,epstopdf,graphicx, amsthm, amsmath, amssymb, times, braket, color, bm}
\usepackage{caption3}
\usepackage{float}
\usepackage[caption = false]{subfig}
\usepackage{cleveref}
\usepackage{makecell}

\definecolor{myurlcolor}{rgb}{0,0,0.7}

\theoremstyle{plain}

\providecommand{\theoremname}{Theorem}
\newcommand*{\myproofname}{Proof}

\makeatother

\theoremstyle{definition}

\theoremstyle{remark}

\usepackage{amscd}

\newcommand{\beq}{\begin{equation}}
\newcommand{\eeq}{\end{equation}}
\newcommand{\ba}{\begin{array}}
\newcommand{\ea}{\end{array}}
\newcommand{\bea}{\begin{eqnarray}}
\newcommand{\eea}{\end{eqnarray}}

\begin{document}

\title{Quantum algorithms for calculating determinant and
inverse of matrix \\ and solving linear algebraic systems}

\author{Alexander I. Zenchuk}
\email{zenchuk@itp.ac.ru}
\affiliation{Federal Research Center of Problems of Chemical Physics and Medicinal Chemistry RAS, Chernogolovka, Moscow reg., 142432, Russia}
\author{Georgii A. Bochkin}
\email{bochkin.g@yandex.ru}
\affiliation{Federal Research Center of Problems of Chemical Physics and Medicinal Chemistry RAS, Chernogolovka, Moscow reg., 142432, Russia}
\author{Wentao Qi}
\email{qiwt5@mail2.sysu.edu.cn}
\affiliation{Institute of Quantum Computing and Computer Theory, School of
Computer Science and Engineering, Sun Yat-sen University, Guangzhou 510006, China}
\author{Asutosh Kumar}
 \email{asutoshk.phys@gmail.com}
 \affiliation{Department of Physics, KSS College, Lakhisarai 811311, India}
\affiliation{P.G. Department of Physics,  Munger University, Munger 811201, India}
\author{Junde Wu}
\email{wjd@zju.edu.cn}
\affiliation{School of Mathematical Sciences, Zhejiang University, Hangzhou 310027, PR~China}

\begin{abstract}

We propose quantum algorithms, purely quantum in nature, for calculating the determinant and inverse of an $(N-1)\times (N-1)$ matrix (depth is $O(N^2\log N)$) which is a simple modification of the algorithm for calculating the determinant of an $N\times N$ matrix (depth is $O(N\log^2 N)$.
The basic idea is to encode each row of the matrix into a pure state of some quantum system.
In addition, we use the representation of the elements of the inverse matrix in terms of algebraic complements.   This algorithm together with that for matrix multiplication { proposed earlier} yields the algorithm for solving systems of linear algebraic equations (depth is $O(N\log^2 N)$. Measurement of the ancilla state with output 1 (probability is $\sim 2^{-O(N\log N)}$) removes the garbage acquired  during calculation.
Appropriate circuits for all three algorithms are presented and have the same estimation $O(N\log N)$ for the space (number of qubits in the circuit).

\end{abstract}
\maketitle

\section{Introduction}

Quantum algorithms exploiting principles of quantum physics can be run on quantum computers and promise to offer considerable advantages over classical counterparts.
Algorithms for quantum calculations are an increasingly prominent field in quantum information research. Notable examples include Shor's factoring algorithm \cite{Sho1,Sho2}, Deutsch parallelism \cite{Deu}, Grover's search algorithm \cite{Gro}, Quantum Fourier transform \cite{QFT1,QFT2,QFT3} and Quantum phase estimation \cite{Wang,QFT3}.
Further advancements include the HHL algorithm for solving systems of linear equations \cite{HHL,HHL1,HHL2,HHL3,HHL4,HHL5,HHL6,HHL7},
in which the crucial point is the conditional rotation of the state of the one-qubit ancilla. This operation, where  inversion of the eigenvalue   requires involving classical computing, is placed between the blocks of  the direct and  inverse phase estimation algorithms.
 HHL algorithm  exponentiates a Hermitian matrix of the considered linear system,   which  can be performed via  the Trotterization formula  \cite{T,AT}.
There is extensive literature devoted to  improvement of this algorithm, especially its scaling and complexity \cite{BHMT,CKS,GSLW,LC,LYC,AL,JRS,SSO,AlL,TAWL}.
 {However, the principal problem of  eigenvalue inversion is not  resolved via  quantum methods yet. For this purpose, for instance, the  quantum state tomography can be used to obtain the eigenvalues  of the exponent of Hermitian operator (the matrix of the linear system)  \cite{T}.}
Nevertheless, the  HHL algorithm  combining the  phase estimation and classical inversion of eigenvalues  is very popular in implementation of  various algorithms of quantum computation and it is widely applicable in quantum machine learning \cite{A,HHL2,RML}.
In particular, it is  applied in the least-square linear-regression algorithms  \cite{WBL,SSP,Wang2} working with large data sets, in quantum algorithms for singular-value decomposition \cite{GSLW,G},  in algorithms for solving linear \cite{B} and nonlinear \cite{LKKLTC} differential equations.

Another method of matrix inversion  is presented in \cite{MRTC, NJ}. Both algorithms are based on the singular value decomposition and apply the function evaluation problem \cite{GSLW}  to approximate the inverse of singular values by odd-polynomial, which requires introducing special scale parameter whose value depends on the length of the interval including nonzero  eigenvalues.

We also refer to algorithms for matrix manipulations based on the Trotterization method and the Baker-Champbell-Hausdorff formula for exponentiating matrices \cite{ZhaoL1}.
{In \cite{GSLW}, algebraic manipulations with matrices are implemented using unitary transformations that include the matrices as specific blocks (block-encoding model). A similar block-encoding approach was used in the algorithm of \cite{DFZ_2020} to embed the inverse of the matrix of an algebraic system into the unitary transformation.
{In \cite{LWWZ}, an algorithm for matrix multiplication using binary encoding  of matrix elements in the states of quantum systems followed by binary multiplication using Toffoli gates was proposed. Matrix multiplication over rings, including the multiplication of Boolean matrices was studied in \cite{KN}.}
}

In our recent work \cite{QZKW_arxive2022}, we presented alternative algorithms for matrix manipulations, such as sum and product of two  matrices, calculation of determinant, matrix inversion, and solving linear systems, based on special unitary transformations of a quantum system.
The main distinction of these algorithms from those mentioned above is the encoding of matrix elements into the pure states of specific (sub)systems as the probability amplitudes. However, the realization of those unitary operators in terms of the basic quantum operations was not explored in that study.

Later, in \cite{ZQKW_arxiv2023}, we proposed the implementation of unitary transformations corresponding to vector inner product, matrix sum, and product using multi-qubit conditional operators (reducible to Toffoli gates) and Hadamard transformations, including the ancilla measurement for garbage removal. The depth  of these algorithms increases logarithmically with the matrix dimension, or even remains independent of it (in the case of matrix sum), demonstrating the advantage of these algorithms compared to their classical counterparts.

In this paper, we  propose quantum algorithms for determinant and inversion of matrix with their application to solving systems of linear equations.
The principal part of {these algorithms} is the algorithm for computing the determinant of a square matrix. It is  built on the ideas of  Refs. \cite{QZKW_arxive2022,ZQKW_arxiv2023}:
\begin{itemize}
\item[-]
Encode matrix elements into the states of certain quantum subsystems through probability amplitudes in such a way that all necessary multiplications of matrix elements are performed automatically and appear in the probability amplitudes of the initial state of the whole system;
\item[-]
Use multiqubit control-SWAP and control-Z operators together with ancilla to organize proper signs of each term in the superposition state and collect them in certain state of the considered system;
\item[-]
Use ancilla measurement to remove the garbage and present the desirable result as a pure state of a certain quantum system. Normalization of this state is defined at the stage of ancilla measurement.
\end{itemize}
We follow the above steps to propose an algorithm for calculating the determinant of a square matrix whose $N$ rows are encoded into the pure states of $N$ separate quantum subsystems. {Notice that creating such initial states is a special problem in itself, but we do not  discuss this problem here.}  The state of the whole system results in all possible products of $N$ elements from different rows.
One only needs to select the proper products to calculate the determinant.
We propose such an algorithm and construct the quantum circuit realizing it.
While developing the algorithm, our focus is on the space and depths of its specific blocks. {Essentially, our algorithm is a quantum realization of the cofactor expansion of the classical algorithm, whose depth exponentially increases with matrix dimension. However, the resulting depth of our algorithm is $O(N\log^2 N)$ (or $O(N^2 \log N)$ in the worst case) with space of $O(N\log N)$ qubits.
{The final step of this algorithm is the one-qubit  ancilla measurement which removes all the garbage. The probability of obtaining the required result of the ancilla measurement (which is 1) diminishes exponentially as the matrix dimension $N$ increases, necessitating an exponentially larger number of algorithm evaluations. This is a central problem that we plan to investigate in future work.}

It is remarkable that the
minor modification of  the algorithm for calculating the determinant yields an alternative quantum algorithm for matrix inversion using its definition in terms of algebraic complements.
We again use the row-wise encoding of  the matrix elements into the state of a quantum system  so that all necessary products of matrix elements appear automatically in the superposition state.   One thus only needs to properly sort these products. The depth of this algorithm is $O(N^2 \log N)$ (or $O(N^2 \log^2 N)$ in the worst case) and the space is $O(N\log N)$ qubits. Based on such inversion algorithm we  develop an alternative algorithm for solving systems of linear algebraic equations. {Thus, we  do not  use Trotterization methods  and classical  subroutine for  inversion of eigenvalues.   Therefore, the proposed algorithm is purely quantum in nature and it  gives an alternative avenue to the  inversion algorithm considered in \cite{HHL,T,TAWL}. The estimations for the depth and space  of this algorithm remain the same.

We also emphasize that, in comparison with inversion algorithms in \cite{MRTC, NJ}, which  approximately calculate  the inverse
eigenvalues, our algorithm constructs exact (up to the available accuracy of realization of quantum operations) inverse matrix without appealing to its  eigenvalues which provides certain advantages of the matrix-encoding  approach.

Regarding the question of algorithmic complexity,  we remark that the specialized classical techniques of matrix manipulations facilitate a reduction in the depth of the algorithms for determinant calculation to $O(\log^2 N)$ \cite{Ber,BAEM} with the size  $O(N^{2.496})$.
In our study, we employ the Leibniz formula as a definition for the matrix determinant and define the inverse matrix in terms of the algebraic complements. Our quantum algorithms based on these definitions demonstrate clear advantages over their classical counterparts. It is plausible that the adaptation of specific classical algorithms for matrix operations into their quantum equivalents could yield further enhancements in the efficiency and complexity of quantum algorithms designed for calculating determinants and inverse matrices. However, this particular aspect is not addressed in the present paper. The primary contribution of our work lies {in promoting} the representation of matrix elements as probability amplitudes of pure states, and we elucidate the benefits of this representation in relation to standard algorithms for matrix manipulations.

The paper is organized as follows. The algorithm for calculating the matrix determinant is proposed in Sec.\ref{Section:det}.
This algorithm is taken as a basis for the algorithm calculating  the inverse matrix  presented in Sec.\ref{Section:inv}. Based on those results, the algorithm for solving systems of linear  algebraic equations is proposed in Sec.\ref{Section:LinSyst}. The main   results are discussed in Sec.\ref{Section:conclusions}. The special subroutine for the controlled measurement is presented in Sec.\ref{Section:CM}.
Appendix, Sec.\ref{Section:appendix}, contains the modified algorithms for matrix multiplication.
and examples of realization of the algorithms  developed in this paper.

\section{Determinant}
\label{Section:det}
To calculate the determinant  of $N\times N$  matrix $\tilde A$ (we assume $N=2^n$)
with elements $a_{jk}$ ($j,k=0,1,\cdots,N-1$),
\begin{eqnarray}\label{AA}
\tilde A=\left(
\begin{array}{cccc}
a_{00}&a_{01}&\cdots& a_{0(N-1)}\cr
a_{10}&a_{11}&\cdots& a_{1(N-1)}\cr
\cdots&\cdots&\cdots&\cdots\cr
a_{(N-2)0}&a_{(N-2)1}&\cdots& a_{(N-2)(N-1)}\cr
a_{(N-1)0}& a_{(N-1)1}&\cdots&a_{(N-1)(N-1)}
\end{array}
\right),
\end{eqnarray}
 we encode each row consisting of $N$ elements into the pure state $|\Psi_j\rangle$ of the $n$-qubit subsystem $S_j$,
\begin{eqnarray}\label{Psi}\label{aa}
&&
|\Psi_j\rangle= \sum_{k=0}^{N-1}a_{jk} |k\rangle_{S_j} ,\;\;j=0,1,\dots,N-1,
\end{eqnarray}
 where $k$ in the state $|k\rangle_{S_j}$ is the binary encoding of the integer $k$, therefore all states $|k\rangle_{S_j}$with fixed $j$ are mutually orthogonal. We shall note that encoding the elements of the matrix into the probability amplitudes of  a quantum state is not a simple problem, but we do not consider aspects of creating such states in this paper.
Representing  rows of the  matrix $\tilde A$ in the form of pure states (\ref{aa})  creates $N$ obvious constraints for the  matrix elements $a_{ij}$ because of the normalization of pure states:
\begin{eqnarray}\label{PsiNorm}
\sum_{k=0}^{N-1}|a_{jk}|^2=1,\;\;j=0,\dots,N-1.
\end{eqnarray}
These constraints can be circumvented by incorporating additional terms with different probability amplitudes into the states $|\Psi_i\rangle$, but this issue will be addressed in future study.
The whole system is the union of the subsystems,
$\displaystyle S=\bigcup_j S_j$,
and the pure state of the whole system reads
\begin{eqnarray}\label{kk}\label{Phi0}
|\Phi_0\rangle &=& |\Psi_0\rangle \cdots  |\Psi_{N-1}\rangle \nonumber \\
&=&
 \sum_{k_0,\dots, k_{N-1}=0}^{N-1} a_{0 k_0} \dots a_{(N-1)k_{N-1}} \nonumber \\
&\times &  |k_0\rangle_{S_{0}}\cdots   |k_{N-1}\rangle_{S_{N-1}} ,
\end{eqnarray}
which includes $N^{(S)}=n N$ qubits.
In computing the determinant, we need only those terms in which all $k_j$ are different. We select these terms from the  whole  pure state of our system:
\begin{eqnarray}\label{kk2}\label{Phi01}
|\Phi_0\rangle &=&
 \sum_{{k_0,\dots, k_{N-1}=0}\atop{k_0\neq\dots \neq k_{N-1}}}^{N-1}  a_{0k_0} \dots a_{(N-1)k_{N-1}} \nonumber \\
&\times &  |k_0\rangle_{S_{0}}
  \cdots   |k_{N-1}\rangle_{S_{N-1}} + |g_1\rangle_S,
\end{eqnarray}
where $|g_1\rangle_S$ collects all extra terms which form the garbage to be removed later. We emphasize that selection of the needed terms in (\ref{Phi01}) is quite formal procedure, at this stage we don't do anything to label these terms.  We will do this later introducing  the ancillae  $B$ and removing  all the garbage via the ancilla measurement, see eqs.(\ref{W2}), (\ref{Phi3D}) and  Fig. \ref{Fig:circuit} .
We have to arrange the appropriate signs for each term of the first part in the right-hand side of (\ref{kk2}). This  can be achieved by setting the subsystem $S_j$ (in each term)  into the state $|j\rangle$ using the SWAP operators together with sign-switching. This pair of operators  performs permutation   and changes the sign of the term.

Thus, our purpose is to exchange the states of subsystems $S_j$ ($j=0,\dots,N-1$) so that the state of the subsystem $S_k$ becomes $|k\rangle_{S_k}$ in each term of the first part of superposition state (\ref{kk2}).
For that we introduce the projectors $P_{S_j}^{(k)}$,
\begin{eqnarray}\label{PS}
&&
P_{S_j}^{(k)} = |k\rangle_{S_j} \; _{S_j}\langle k|,\\\nonumber
&& k=0,\dots  N-2, \;\;j=k+1,\dots  N-1,
\end{eqnarray}
 and
the $\tilde n_k$-qubit  ancillae $A_k$, ($\tilde n_k = \lceil \log(N-k) \rceil$,  $k=0,\dots  N-2$).  The ancilla $A_k$ serves to save the index $j$ of the subsystem $S_j$
whose state is $|k\rangle$ with $j>k$. Therefore, the number of qubits $\tilde n_k$ in $A_k$ is less than $n=\log N$ unless $k=0,1$: $\tilde n_k\le n$.

For instance,
in Fig. \ref{Fig:circuit} for the case $N=4$,  there are three ancillae $A_0, A_1, A_2$.  Ancilla $A_2$ has to encode either only one state  $|2\rangle_{S_3}$ or nothing, therefore one qubit is
enough: we can encode $|2\rangle_{S_3}$ by the state $|1\rangle_{A_2}$ of ancilla $A_2$. Next, $A_1$ has to encode two states $|1\rangle_{S_2}$ and $|1\rangle_{S_3}$ or nothing. Two qubits are enough for that:
$|1\rangle_{S_2}\rightarrow |1\rangle_{A_1}$,  $|1\rangle_{S_3}\rightarrow |2\rangle_{A_1}$. Similarly, $A_0$ requires 2 qubits as well: $|0\rangle_{S_1}\rightarrow |1\rangle_{A_0}$, $|0\rangle_{S_2}\rightarrow |2\rangle_{A_0}$, $|0\rangle_{S_3}\rightarrow |3\rangle_{A_0}$.  Thus, we have the map
\begin{eqnarray}\label{map}
|k\rangle_{S_j} \to |\tilde j\rangle_{A_k},
\end{eqnarray}
which can be realized as follows.

Using the projector $P_{S_j}^{(k)}$ and ancilla $A_k$ we can prepare
the operator $V_{S_jA_k}$, which relates  the index $j$ of $S_j$ to the appropriate state $|\tilde j\rangle_{A_k}$ of the ancilla $A_k$:
\begin{eqnarray}\label{Vn}
V_{S_jA_k} &=& P_{S_j}^{(k)}\otimes Z_k(j)\sigma^{(x)}_{kj} +
(I_{S_j} -  P_{S_j}^{(k)}) \otimes I_{A_k},
\end{eqnarray}
where
$\sigma^{(x)}_{kj}$ includes the set of $\sigma^{(x)}$-operators which change the ground state of the ancilla $A_k$ to the state $|\tilde j\rangle_{A_k}$, subscripts $k$ and $j$ are  related to $k$ and $j$ in the map (\ref{map}). For instance, to map the state $|0\rangle_{S_2}$ into the state $|2\rangle_{A_0}$ of the two-qubit ancilla $A_0$ we need to apply $\sigma^{(x)}$ to the second qubits of $A_0$, i.e.,  $\sigma^{(x)}_{02} = I \otimes \sigma^{(x)}$, where $I$ is the identity operator applied to the first qubit of $A_0$;
$Z_k(j)$ changes the sign of the ancilla state
{$ |\tilde j\rangle_{A_{k}}$ (essentially, it counts swaps and  can be represented by either the  $\sigma^{(z)}$-operator applied to one of the excited qubits, or by the $\sigma^{(x)}\sigma^{(z)}\sigma^{(x)}$-operator applied to one of the ground-state  qubits} (only to one!) of $A_k$).

 It is hard to derive the explicit formula for $Z_k(j)$ in general case, but it can be simply done in any particular case. For instance, in the above example of the transformation $|0\rangle_{S_2} \to |2\rangle_{A_0}$ we can set $Z_0(2) =I\otimes  \sigma^{(z)}$. Thus, the operator $V_{S_jA_k}$ acts as follows. If the state of $S_j$ is $|k\rangle_{S_j}$, then the state of $A_k$ becomes $|\tilde j\rangle_{A_k}$, otherwise the state of $A_k$ remains $|0 \rangle_{A_k}$.

To arrange the exchange of states between $S_j$ and $S_k$, we introduce
the projectors $P_{A_k}^{(j)}$,
\begin{eqnarray}\label{PA}
&&
P_{A_k}^{(j)} = |\tilde j\rangle_{A_k} \; _{A_k}\langle \tilde j|,\;\;
k=0,\dots  N-2,\\\nonumber
&&j=k+1,\dots  N-1,\;\;\tilde j=0,\dots  N-k-2, \;\;\tilde j =j-k-1,
\end{eqnarray}
and construct the operators $U_{S_kS_j}$:
\begin{eqnarray}\label{U}
U_{S_k S_{j}} &=& \Big(
SWAP_{S_k,S_{j}} \otimes
P_{A_k}^{(j)}  \nonumber \\
 &+& I_{S_kS_{j}}  \otimes (I_{A_k} - P_{A_k}^{(j)})\Big) ,\;\;k\neq j .
\end{eqnarray}
Here, the operator $SWAP_{S_k,S_{j}}$ swaps the states of the subsystems $S_k$  and $S_{j}$ by swapping states of appropriate qubits of these subsystems. Thus, the operator $U_{S_k S_{j}}$  swaps the states of $S_k$ and $S_j$ if only he state of $A_k$ is $|\tilde j \rangle_{A_k}$. Otherwise the states of the above subsystems remain unchanged.

The circuits for the operators $V_{S_jA_k}$ and
$U_{S_k S_{j}}^{(k)}$ are given in Fig.\ref{Fig:circuitH}(a) (the general matrix case) and in Fig.\ref{Fig:circuitH}(b) (the particular $4\times 4$-matrix case).
\begin{figure*}[!]
\centering
\includegraphics[width=1\textwidth,angle=0]{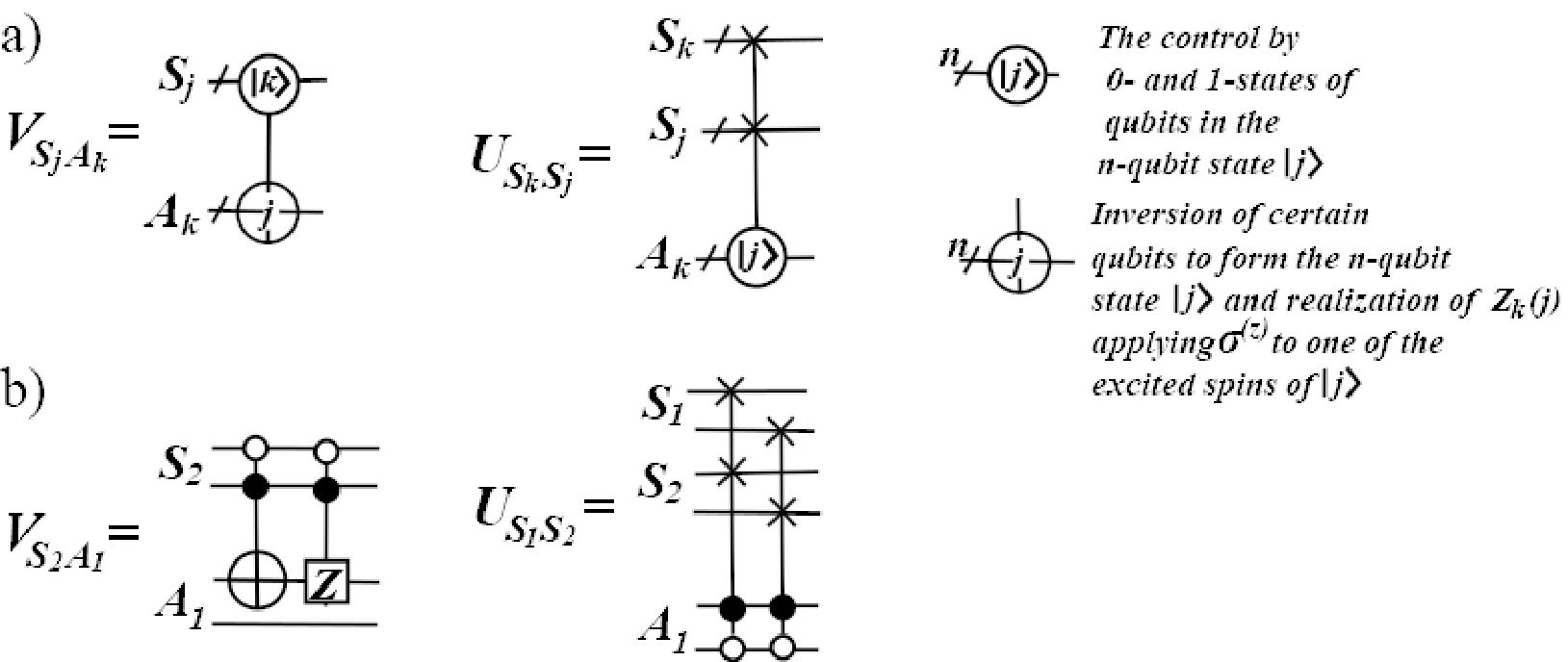}
\caption{$V$ and $U$ operators for (a) the general case and (b) the  $4\times 4$-matrix case. In the latter case, $\tilde n_0=\tilde n_1=2$, $\tilde n_2=1$, $\tilde n_k < n$. Hence, $A_1$ is a two-qubit ancilla.}
\label{Fig:circuitH}
\end{figure*}
The depth of each operator $V_{S_jA_k}$ is $O(n)$ (due to the projectors $P_{S_j}^{(k)}$ with $n$ control qubits and we assume that all $\sigma^{(x)}$-operators of $\sigma^{(x)}_{kj}$ are applied in parallel to different qubits of  $A_k$), while the depth of  $U_{S_k S_{j}}$ is the depth of the control-SWAP-operator which includes  $n$ swaps each controlled by $n$ qubits, i.e., the total depth  is $n O(n)=O(n^2)$. Theoretically, it can be implemented with less operations but this would not change the asymptotic complexity of the circuit.

{Applying the operators
\begin{eqnarray}\label{W}
W_{k} = \prod_{j=k+1}^{N-1} U_{S_k S_{j}} V_{S_jA_k},\;\; k=0,\dots,N-2,
\end{eqnarray}
to the state $|\Phi_0\rangle |0\rangle_A$, we put the subsystem $S_k$ into the  state $|k\rangle_{S_k}$ in each term of  the first part of  (\ref{kk2}).}
Although $W_k$ has product of $N-k-1$ pairs of $U$- and $V$-operators, only one pair of such operators is applied to each particular term in $|\Phi_0\rangle$ which is due to the structure of the projector operators $P^{(k)}_{S_j}$. We assume that this means the $UV$-products can be applied in parallel and thus the depth of $W_k$ is $O(n)+ O(n^2) = O(n^2)$. Otherwise we have to multiply the obtained depth by $O(N)$.
 The operator $\prod_{k=0}^{N-2} W_k$ provides the needed sign of the probability amplitudes which will compose the expression for the determinant. To collect all of them in a single probability amplitude  we apply the Hadamard transformation to each qubit of ancillae  $A_k$, $k=0,\dots,N-2$, and  denote this transformations  $H_A$, i.e.
\begin{eqnarray}\label{HA}
H_A= H_{A_0} \dots H_{A_{N-2}},\;\; H_{A_k} = \bigotimes_{k=0}^{\tilde n_k-1} H.
\end{eqnarray}}
There are $ \tilde N$ Hadamard operators in $H_A$,
\begin{eqnarray}\label{tN}
 \tilde N=\sum_{k=0}^{N-2} \tilde n_k = O(N\log N).
\end{eqnarray}
 Applying the operator $W^{(1)}_{SA}$,
\begin{eqnarray}
\label{W1}
W^{(1)}_{SA}=H_A \prod_{k=0}^{N-2} W_k,
\end{eqnarray}
 to the state $|\Phi_0\rangle |0\rangle_A$,
$A=\bigcup_{k=0}^{N-2}A_k$, and selecting the terms which carry the information about the determinant (these terms correspond to the ground state of $A$)  we obtain
\begin{eqnarray}
&&
|\Phi_1^{(D)}\rangle =W^{(1)}_{SA} |\Phi_0\rangle |0\rangle_A \\\nonumber
&&=
 \frac{\det(\tilde A)}{2^{\tilde  N/2}}\;
 |0\rangle_{S_0}|1\rangle_{S_1} \dots  |{N-1}\rangle_{S_{N-1}}
|0\rangle_{A}+ |g_2\rangle_{SA}.
\end{eqnarray}
Here we introduce the determinant $\det(\tilde A)$ represented by  the Leibniz formula,
\begin{eqnarray}\label{det}
\det(\tilde A) &=&  \sum_{k_0\neq\dots\neq k_{N-1}} \epsilon_{k_0\dots k_{N-1}} a_{ k_00}\dots  a_{k_{N-1}(N-1)},
\end{eqnarray}
where
$\epsilon_{k_0\dots k_{N-1}} $ is the permutation tensor: 
\begin{eqnarray}\nonumber
\epsilon_{k_0\dots k_{N-1}}  =\left\{
\begin{array}{ll}
(-1)^{{\mbox{perm}}(k_0\dots k_{N-1})},&k_0 \neq k_1 \neq \dots\neq  k_{N-1}\cr
0& {\mbox{otherwise}}
\end{array}
\right..
\end{eqnarray}
 The depth  of $W^{(1)}$ is $O(N n^2)$.
Next, we can set  states of all $S_j$ in the first term of $|\Phi_1^{(D)}\rangle$ into the $|0\rangle_{S_j}$ states. To this end we apply the $\sigma^{x}$-operators to the excited qubits of $S$, i.e. prepare the operator  $X_S$.
 \begin{eqnarray}
\label{XS}
X_S=\prod_{j=1}^{N-1} X_{S_j}(j),
\end{eqnarray}
 with $X_{S_j}(j)$
 being the set of $\sigma^{(x)}$ operators acting on
  the excited qubits of $S_j$ in the state $|j\rangle_{S_j}$.   The depth of $X_S$  is $O(1)$.  Applying $X_S$ to $|\Phi_1^{(D)}\rangle$ we obtain:
\begin{eqnarray}
&&|\Phi_2^{(D)}\rangle= X_S |\Phi_1^{(D)}\rangle \\\nonumber
 &&=
\frac{\det(\tilde A)}{2^{\tilde N/2}}\;
 |0\rangle_{S}
|0\rangle_{A}+ |g_3\rangle_{SA} .
\end{eqnarray}
We note that the operators $H_A$ in  $W^{(1)}_{SA}$ and operator $X_S$ can be applied simultaneously because they affect  different qubits. We separate them to underline the distinction in their  functions.
Next, we remove the garbage. For that we introduce the projector
$P_{SA}=|0\rangle_{S} |0\rangle_{A}\; _{A}\langle 0|  _{S}\langle 0|$, include the one-qubit ancilla $B$ in the ground state
and construct the operator $W^{(2)}_{SAB}$,
\begin{eqnarray}
\label{W2}
W^{(2)}_{SAB}=P_{SA} \sigma^{(x)}_{B} + (I_{SA}-P_{SA}) I_{B},
\end{eqnarray}
 whose depth is $O(Nn +Nn) = O(Nn)$.
 Applying it to $|\Phi_2^{(D)}\rangle |0\rangle_B$ we obtain
\begin{eqnarray}\label{Phi3D}
&&|\Phi_3^{(D)}\rangle= W^{(2)}_{SAB} |\Phi_2^{(D)}\rangle|0\rangle_B \\\nonumber
&&=
 \frac{\det(\tilde A)}{2^{\tilde N/2}}\;
 |0\rangle_{S}|0\rangle_{A}
|1\rangle_{B}+
 |g_3\rangle_{SA}|0\rangle_B .
\end{eqnarray}

Now we measure the ancilla $B$ with the output $|1\rangle_{B}$. The probability of this result of  measurement is $2^{-\tilde N} |{\det(\tilde A)}|^2$. After the measurement we get the following state:
\begin{eqnarray}\label{Phi4D}
&&|\Phi_4^{(D)}\rangle= |\Psi^{(D)}_{out}\rangle |0\rangle_{A}, \;\;  |\Psi^{(D)}_{out}\rangle =   \arg(\det(\tilde A)) \;
 |0\rangle_{S} , 
\end{eqnarray}

The particular circuit for calculating the $4\times 4$ determinant using the proposed algorithm is given in Fig. \ref{Fig:circuit}.  The circuit for the determinant of $N\times N$ matrix includes  $N\log N$ qubits of the subsystem $S$, $\tilde N = O(N\log N)$ qubits of the subsystem $A$ and one qubit of the ancilla $B$, therefore the space of this circuit is $O(N\log N)$ qubits.

\begin{figure*}[!]
\noindent
\includegraphics[width=1.\textwidth,angle=0]{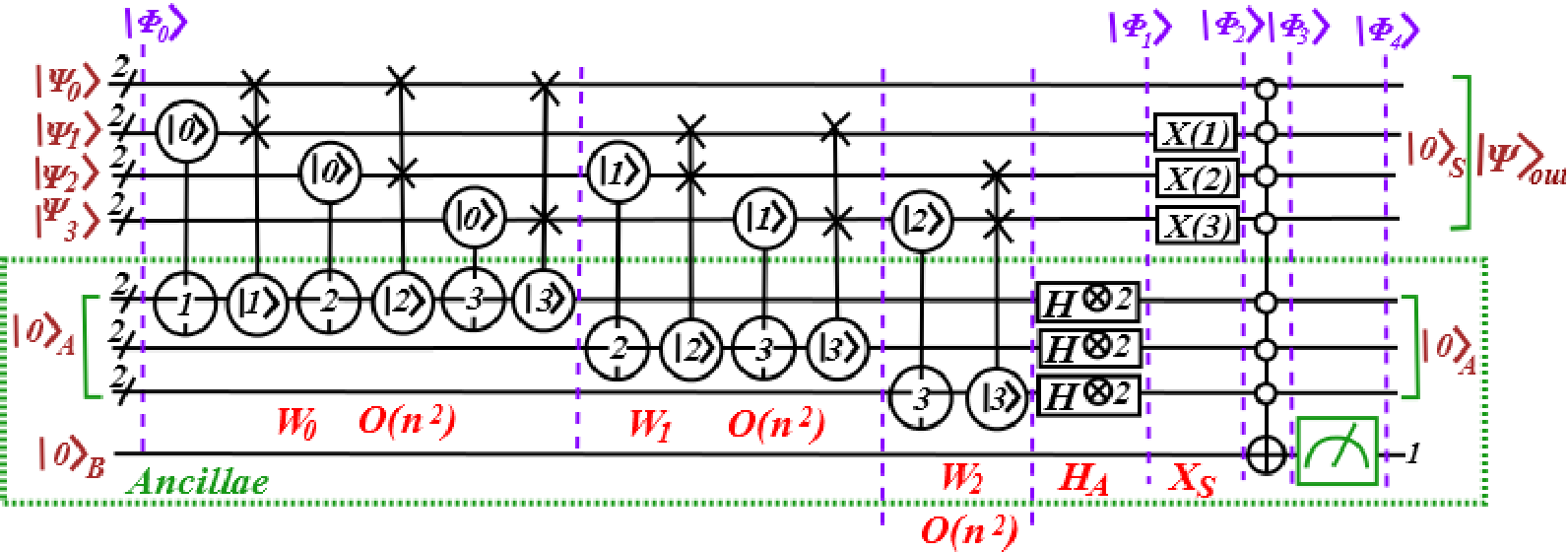}
\newline
\includegraphics[width=0.3\textwidth,angle=0]{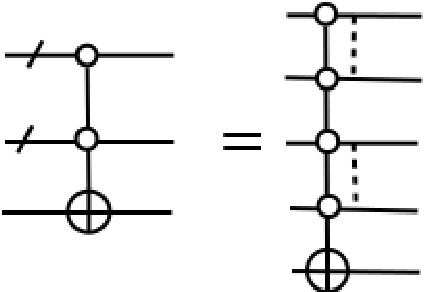}
\caption{Quantum circuit illustrating computation of determinant of a $4\times 4$ matrix, see notations in Fig.\ref{Fig:circuitH}. We omit superscript $(D)$ in $|\Phi_k\rangle$ $(k=1,\dots,4)$ and $|\Psi_{out}\rangle$ for brevity. Operators $H_A$ and $X_S$ can be applied in parallel since they affect different qubits. The lower circuit is notation for multi-qubit controlled $\sigma^{(x)}$ operation.  }
\label{Fig:circuit}
\end{figure*}

\section{Inverse matrix}
\label{Section:inv}
We remark here that the minor modification of the algorithm for calculating the determinant of a square $N\times N$ matrix  in Sec. \ref{Section:det} allows to calculate the inverse of a square $(N-1)\times (N-1)$ matrix.  Moreover, this modification leads to a minor  increase in the depth of the algorithm which becomes  $O(N^2\log N)$ instead of $O(N\log^2 N)$ in the algorithm for calculating the determinant. These facts motivate working out  the quantum inversion algorithm based on the definition of the inverse matrix in terms of  algebraic complements, although quantum analogues of other algorithms of matrix inversion might be  more promising. 

\subsection{Preliminary consideration}
\label{Section:pa}

We impose a special structure on the elements in the first row and first column of the matrix $\tilde A$ given in (\ref{AA}):
\begin{eqnarray}\label{cconst}
a_{i0}=q, \;\;a_{0 j} =1/\sqrt{N}, \;\;i,j=0,\dots,N-1,
\end{eqnarray}
i.e.,
\begin{eqnarray}\label{tA}
\tilde A=\left(
\begin{array}{cccc}
\frac{1}{\sqrt{N}}&\frac{1}{\sqrt{N}}&\cdots&\frac{1}{\sqrt{N}}\cr
q&a_{11}&\cdots& a_{1(N-1)}\cr
q&a_{21}&\cdots& a_{2(N-1)}\cr
\cdots&\cdots&\cdots&\cdots\cr
q&a_{(N-1)1}&\cdots& a_{(N-1)(N-1)}
\end{array}
\right).
\end{eqnarray}
The algorithm discussed below allows to invert the $(N-1)\times (N-1)$ matrix $A$ {(rather then the matrix $\tilde A$ itself)} which is embedded into $\tilde A$,
\begin{eqnarray}
A=\left(
\begin{array}{ccc}
a_{11}&\cdots& a_{1(N-1)}\cr
a_{21}&\cdots& a_{2(N-1)}\cr
\cdots&\cdots&\cdots\cr
a_{(N-1)1}&\cdots& a_{(N-1)(N-1)}
\end{array}
\right).
\end{eqnarray}

According to the standard definition,
the element $A^{-1}_{ji}$ of the inverse matrix is expressed in terms of the minor $M_{ij}$ obtained by removing the $i$th row and $j$th column of $A$:
\begin{eqnarray}
A^{-1}_{ji} =\frac{ (-1)^{i+j} M_{ij}}{\det(A)} .
\end{eqnarray}
From another point of view,  the minor $M_{ij}$ appears among the terms included in $\det(\tilde A)$. In fact, one of the terms in this determinant is following (remove the $(i+1)$th row and 1st column from $\tilde A$; notice that $(i+1)$th row contains the elements $q$, $a_{i1}$, $a_{i3}$, ..., $a_{i(N-1)}$):
\begin{eqnarray}\label{det1}
\frac{(-1)^{i+2} q }{\sqrt{N}}
\left|
\begin{array}{ccc}
1&\cdots&1\cr
a_{11}&\cdots& a_{1(N-1)}\cr
\cdots&\cdots&\cdots\cr
a_{(i-1)1}&\cdots& a_{(i-1)(N-1)}\cr
a_{(i+1)1}&\cdots& a_{(i+1)(N-1)}\cr
\cdots&\cdots&\cdots\cr
a_{(N-1)1}&\cdots& a_{(N-1)(N-1)}
\end{array}
\right|,
\end{eqnarray}
In turn, this term includes the following term (remove the 1st row and $j$th column from the determinant (\ref{det1});  the $j$th column contains the elements $1$, $a_{1j}$, $a_{2j}$, ..., $a_{(N-1)j}$):
\begin{widetext}
\begin{eqnarray}\label{det2}
&&-\frac{(-1)^{i+j} q }{\sqrt{N}}
\left|
\begin{array}{cccccc}
a_{11}&\cdots & a_{1(j-1)}& a_{1(j+1)}&\cdots&   a_{1(N-1)}\cr
\cdots&\cdots &\cdots&\cdots&\cdots&\cdots \cr
a_{(i-1)1}&\cdots &a_{(i-1)(j-1)}& a_{(i-1)(j+1)}&\cdots&  a_{(i-1)(N-1)}\cr
a_{(i+1)1}&\cdots &a_{(i+1)(j-1)}& a_{(i+1)(j+1)}&\cdots& a_{(i+1)(N-1)}\cr
\cdots &\cdots&\cdots\cr
a_{(N-1)1}&\cdots&a_{(N-1)(j-1)}& a_{(N-1)(j+1)}&\cdots&     a_{(N-1)(N-1)}
\end{array}
\right| \\ \nonumber
&&
= - \frac{q}{\sqrt{N}} (-1)^{i+j} M_{ij}=-\frac{\det(A) q }{\sqrt{N}} A^{-1}_{ji}.
\end{eqnarray}
\end{widetext}
Thus, each  element $A^{-1}_{ji}$ of the inverse matrix is included into $\det(\tilde A)$ up to a factor which is the same for all elements.
All what remains is to extract all the terms  composing the element $A^{-1}_{ji}$ from those  composing  $\det(\tilde A)$ and encode them into the proper state of some quantum subsystem.
We show that  this can be simply realized.

\subsection{Basic algorithm}
\label{Section:invB}

The circuit for the algorithm  is presented in Fig.\ref{Fig:circuitInv-LSBlock}.
According to (\ref{Psi}),  the pure state $|\Phi_0\rangle$ of the whole system is given in (\ref{kk})
with normalization (\ref{PsiNorm}) and  constraints  (\ref{cconst}) for the  elements in the first row and first column of the matrix $\tilde A$, see (\ref{tA}).
From this state, we select only those terms that compose the determinant of $\tilde A$ and therefore contain the terms that appear in the resulting inverse matrix $A^{-1}$, i.e., we consider $|\Phi_0\rangle$ in form (\ref{kk2}).

\begin{center}
\begin{figure*}[htb]
\subfloat[]{\includegraphics[width=5in, angle=0]{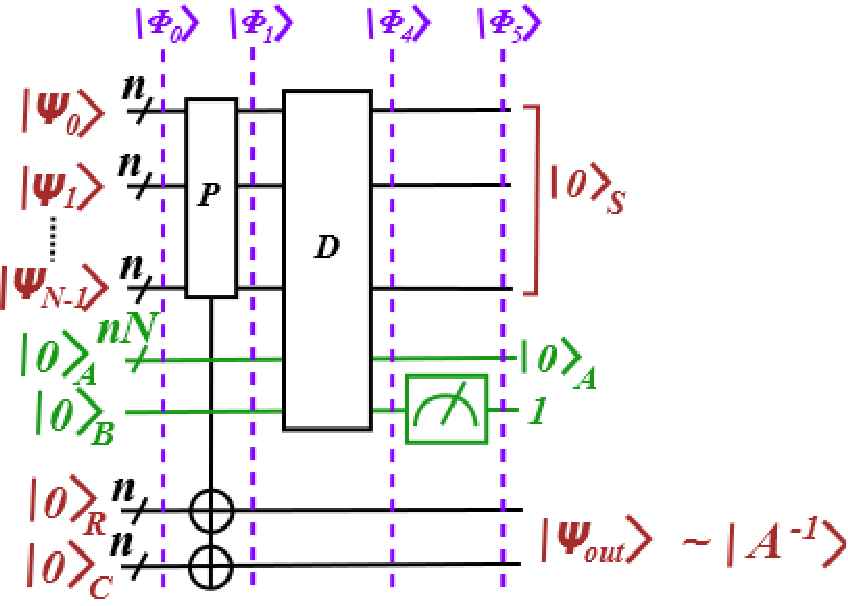}} \hspace{0.3cm}
\subfloat[]{\includegraphics[width=7in, angle=0]{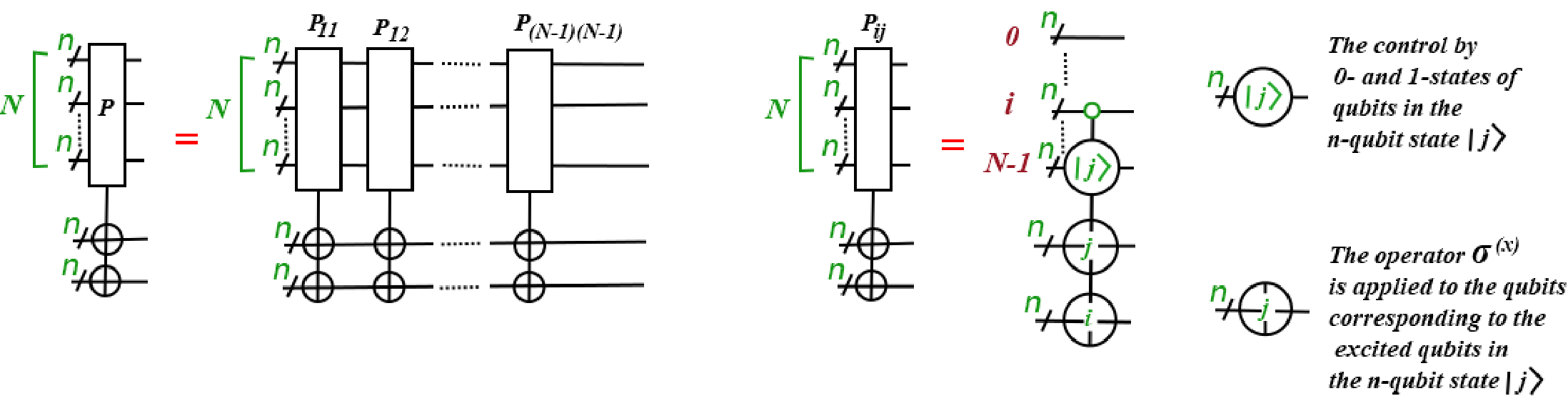}}
\caption{(Color online) (a) Structure of quantum circuit for matrix inversion (ancillae are shown in  green).  (b) The structure of the block $P$}
\label{Fig:circuitInv-LSBlock}
\end{figure*}
\end{center}

However, not all terms in the first part of $|\Phi_0\rangle$ are needed in our algorithm. We have to keep only those of them that include the  1st power of the  factor $q$, (see constraints (\ref{cconst})),  i.e., exactly one  state $|0\rangle_{S_j}$, $1\le j\le N-1$, can appear in  the
product  $|k_1\rangle_{S_{1}} \dots  |k_{N-1}\rangle_{S_{N-1}}$.
We also have to associate each of the   probability amplitudes in (\ref{Phi01})  with the appropriate  element  $A^{-1}_{ij}$ of the inverse matrix.
To do so, we introduce two $n$-qubit subsystems $R$ and $C$ 
 whose basis states serve to enumerate, respectively, rows and columns of the inverse matrix $A^{-1}$.

To select the terms that compose $A^{-1}_{ji}$ from the superposition state  we use the projector
\begin{eqnarray}\label{proj}
P_{ij} = |0\rangle_{S_{i}}\;_{S_{i}}\langle 0|  \otimes  |j\rangle_{S_0}\; _{S_0}\langle j|          , \;\;i, j=1,\dots,N-1. 
\end{eqnarray}
The idea is that this projector allows labeling  the 1st and $(j+1)$th ($j=0,\dots,N-1$) columns and the 1st and  $(i+1)$th ($0\le i\le N-1$) rows in  the matrix $\tilde A$  (\ref{tA}).   As the result, we  select  those terms from the superposition state whose   probability amplitudes compose $M_{ij}$.

Now we can  construct the controlled operator
\begin{eqnarray}
W_{SRC}^{(ij)} = P_{ij}\otimes \sigma^{(x)}_{R^{(j)}C^{(i)}} + (I _S-P_{ij})\otimes I_{RC}.
\end{eqnarray}
This operator conditionally applies  $\sigma^{(x)}_{R^{(j)}C^{(i)}}$ which is the product of $\sigma^{(x)}$ (acting on
the subsystems $R$ and $C$) such that $\sigma^{(x)}_{R^{(j)}C^{(i)}}|0\rangle_R |0\rangle_C = |j\rangle_R |i\rangle_C$.
The depth of $W_{SRC}^{(ij)}$  is $O(n)$. To establish it we use the fact that the controlled operator    $W_{SRC}^{(ij)}$  with $n$-qubit controlling register can be represented in terms of Toffoli gates by a circuit of depth $O(n)$ \cite{KShV} and require that all $\sigma^{(x)}$ operators in $\sigma^{(x)}_{R^{(j)}C^{(i)}}$  can be applied simultaneously.
Applying the operator ${\hat W}_{SRC}=\prod_{i=1}^{N-1}\prod_{j=1}^{N-1}  W_{SRC}^{(ij)}$ (whose depth is $O(N^2n)$)  to $|\Phi_0\rangle  |0\rangle_R |0\rangle_C$ we obtain
\begin{widetext}
\begin{eqnarray}\label{Phi10}
&&
|\Phi_1\rangle = \hat{W}_{SRC}|\Phi_0\rangle  |0\rangle_R |0\rangle_C \\ \nonumber
&&
= \frac{1}{\sqrt{N}}\sum_{{i=1}}^{N-1} \sum_{{j=1}}^{N-1} \sum_{{k_1\dots k_{i-1},k_{i+1}\dots k_{N-1}=0} \atop{{0\neq k_1\neq \dots \neq k_{i-1} \atop{\neq k_{i+1}\neq\dots\neq k_{N-1} \neq j}}}}^{N-1}
a_{1k_1}\dots a_{(i-1) k_{i-1}} a_{i0} a_{(i+1) k_{i+1}} \dots a_{(N-1)k_{N-1}}\times
 \\\nonumber
&&
|j\rangle_{S_0} |k_1\rangle_{S_1} \dots |k_{i-1}\rangle_{S_{i-1}} |0\rangle_{S_{i}}|k_{i+1}\rangle_{S_{i+1}} \dots  |k_{N-1}\rangle_{S_{N-1}}
 |j\rangle_R |i\rangle_C +  |g_2\rangle_{SRC},
\end{eqnarray}
\end{widetext}
where $a_{i0}=q$. In the rhs of  (\ref{Phi10}), the first part collects those terms of the state $|\Phi_1\rangle$ that are labelled by the states $|j\rangle_R$ and $|i\rangle_C$ of the subsystems $R$ and $C$   respectively with $j>0$ and $i>0$. That is why the states of $S_0$ and $S_i$ are, respectively, $|j\rangle_{S_0}$ and $|0\rangle_{S_i}$  in the selected part of $|\Phi_1\rangle$.
Now we have to arrange the proper sign of each term in the first part of   (\ref{Phi10}). According to Sec.\ref{Section:pa},  this can be done via calculating the determinant of $\tilde A$  using the algorithm proposed in Sec.\ref{Section:det}.
Applying the transformation $W^{(1)}_{SA}$ given in (\ref{W1}) to $|\Phi_1\rangle$ we obtain
\begin{eqnarray}\label{PHI2}
&&
| \Phi_2\rangle =W^{(1)}_{SA} |\Phi_1\rangle |0\rangle_A \\\nonumber
&&=-
\frac{q  \det(A)}{2^{(\tilde N+n)/2}}
\sum_{{i=1}}^{N-1} \sum_{{j=1}}^{N-1} A^{-1}_{ji}
 |0\rangle_{S_0}|1\rangle_{S_1}\\ \nonumber
&& \dots \otimes |{N-1}\rangle_{S_{N-1}}|j\rangle_R |i\rangle_C
|0\rangle_{A}+ |g_3\rangle_{SRCA} .
\end{eqnarray}
Next, applying the operator $X_S=\prod_{j=1}^{N-1} X_{S_j}(j)$ given in   (\ref{XS}) to $| \Phi_2\rangle$
we obtain (we note that the  subsystem $S=\cup_{j=0}^{N-1}S_j$ has the same number of qubits in both algorithm of determinant calculation and  algorithm of matrix inversion, therefore we can apply the same operator $X_S$ in both cases)
\begin{eqnarray}\label{PHI3}
&&|\Phi_3\rangle= X_S |\Phi_2\rangle \\\nonumber
 &&=
 -\frac{q  \det(A)}{2^{(\tilde N+n)/2}}\sum_{{i=1}}^{N-1} \sum_{{j=1}}^{N-1} A^{-1}_{ji}
 |0\rangle_{S}  |j\rangle_R |i\rangle_C
|0\rangle_{A}+ |g_4\rangle_{SRCA} ,
\end{eqnarray}
Finally, applying
$W^{(2)}_{SAB}$ (\ref{W2}) to  $|\Phi_3\rangle |0\rangle_B$ we obtain
\begin{eqnarray}\label{Phi2}
&& |\Phi_4\rangle = W^{(2)}_{SAB} |\Phi_3\rangle|0\rangle_B \nonumber \\
&& =\sum_{{i=1}}^{N-1} \sum_{{j=1}}^{N-1} \tilde a_{ji}
|0\rangle_S  |j\rangle_R |i\rangle_C| 0\rangle_A |1\rangle_{B}\nonumber \\
&&+|g_5\rangle_{SRCA}|0\rangle_B,
\end{eqnarray}
where
\begin{equation} \label{ajl}
\tilde a_{ji} = -\frac{q\det(A) A^{-1}_{ji}}{2^{(\tilde N+n)/2}}.
\end{equation}
The depth of the operator $D_{SAB}=W^{(2)}_{SAB}X_SW^{(1)}_{SA}$ (under certain assumptions on the physical realization of the operator $W_k$, see Eq.(\ref{W})) is $O(N n^2)< O(N^2 n)$   (the depth of $\hat W_{SRC}$).
After measuring the ancilla $B$ to get the output $|1\rangle$ with the probability $\frac{q^2  |\det(A)|^2}{2^{\tilde N+n }}  G^2 \sim 2^{-O(\tilde N) }= 2^{-O(Nn)})$, $G=\sqrt{\sum_{i,j=1}^{N-1}|A^{-1}_{ij}|^2 }$,  we get the desired result
\begin{eqnarray}\label{Phi5}
|\Phi_5\rangle = |\Psi_{out}\rangle
|0\rangle_S | 0\rangle_A,
\end{eqnarray}
where
\begin{eqnarray}
 \label{Psiout}
|\Psi_{out}\rangle &=& G^{-1}\sum_{{i=1}}^{N-1} \sum_{{j=1}}^{N-1} A^{-1}_{ji}   |j\rangle_R |i\rangle_C =G^{-1}  |A^{-1}\rangle.
\end{eqnarray}
The state $|\Psi_{out}\rangle$ encodes $A^{-1}$ in  the normalized form, i.e., $\langle\Psi_{out}|\Psi_{out}\rangle =1$, while the state $|A^{-1}\rangle=\sum_{{i=1}}^{N-1} \sum_{{j=1}}^{N-1} A^{-1}_{ji}   |j\rangle_R |i\rangle_C$ is not normalized. The depth of discussed algorithm for matrix inversion is determined by the operator $\hat W_{SRC}$ and equals $O(N^2 \log N)$. The circuit includes $N\log N$ qubits of the subsystem $S$, $\tilde N = O(N\log N)$ qubits of the ancilla $A$, one qubit of the ancilla $B$ and $2 \log N$ qubits of $R$ and $C$, thus the {space} of this circuit is approximated by the {space} of the circuit for calculating the determinant which is $O(N\log N)$.

 We note that this  depth is obtained under assumption that the depth of the operator $W^{(1)}_{SA}$ in (\ref{W1}) is $N\log^2 N$. If this depth is $N^2\log^2 N$, then the depth of the matrix inversion algorithm is also $N^2\log^2 N$.

\section{Solving systems of linear algebraic equations}
\label{Section:LinSyst}
Having constructed the inverse matrix $A^{-1}$ encoded into the state of  the subsystem $R\cup C$, we obtain  a tool for solving a  system of linear algebraic equations represented in the matrix form, $A\boldsymbol{x}=\boldsymbol{b}$, whose solution reads $\boldsymbol{x}= A^{-1}\boldsymbol{b}.$

Therefore, we have to apply the multiplication algorithm proposed in Ref.\cite{ZQKW_arxiv2023} (see also Appendix, Sec.\ref{Section:prod}, for its modified version)
to the matrices $A^{-1}$ and $\boldsymbol{b}$.
The structure of the circuit is shown in Fig.\ref{Fig:LinSyst}(a). More detailed circuit is presented in Fig.\ref{Fig:LinSyst}(b).

\begin{figure*}[!]
\centering
\subfloat[]{\includegraphics[width=4in, angle=0]{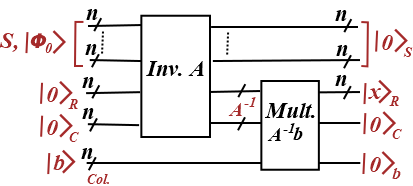}}
\newline
\subfloat[]{\includegraphics[width=0.8\textwidth,angle=0]{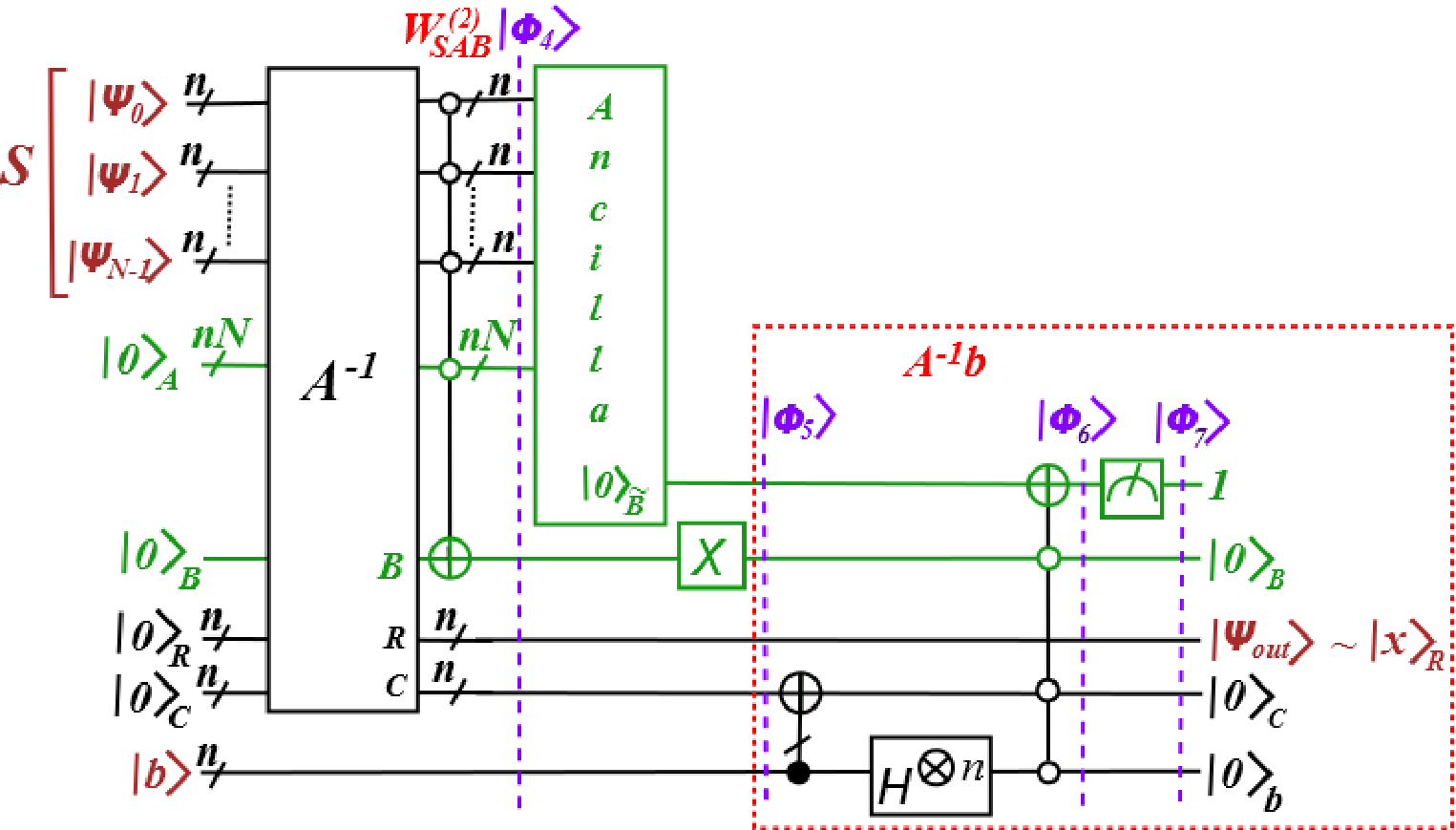}}
\caption{(Color online) (a) Structure of quantum circuit for solving a system of linear algebraic equations. Both blocks share the same ancillae (not shown in figure). (b)  Circuit for quantum algorithm solving a system of linear algebraic equations $A\boldsymbol{x}=\boldsymbol{b}$. Circuit of matrix inversion (up to the operator $W^{(2)}_{SAB}$)  is denoted by a block  $A^{-1}$. All input qubits of the matrix-inversion algorithm (system $S$) become ancillae qubits. The output of the matrix-inversion algorithm (systems $R$ and $C$) together with system $b$ encoding the vector $\boldsymbol{b}$ become the input of the matrix-multiplication block. The output of the whole algorithm is encoded in the state $|x\rangle_R$ of the system $R$. All ancillae qubits are shown in green color. We omit the superscript $(L)$ in $|\Phi_k\rangle$ $(k=5,6,7)$ and $|\Psi_{out}\rangle$ for brevity.}
\label{Fig:LinSyst}
\end{figure*}

Notice that we can use the same ancillae in both blocks of this circuit (ancillae are not shown in Fig.\ref{Fig:LinSyst}(a)). Moreover,  the subsystem $S$ encoding the matrix $A$ can be used as ancilla in further calculations.

In this case we do not measure ancillae $B$ (unlike Sec.\ref{Section:invB}) and thus start with the state $|\Phi_4\rangle$ given in (\ref{Phi2}).
The vector $\boldsymbol{b}$ is encoded as
\begin{eqnarray}
|b\rangle = \sum_{j=0}^{N-1} b_j |j\rangle_b,\;\; b_0=0, \;\;\sum_{j=1}^{N-1} |b_j|^2=1.
\end{eqnarray}
The zero value for $b_0$ is prescribed by the encoding of $A^{-1}$ in (\ref{Psiout}), where the $0$-states of $R$ and $C$ are not used.  Out of system $S\cup A $ (which forms an ancilla in Fig.\ref{Fig:LinSyst}b), we take the one-qubit ancilla $\tilde B$ needed in the multiplication  algorithm  and neglect other qubits. We also apply the operator $X_B =\sigma^{(x)}$ to invert the state of $B$. Thus,
the state of the whole system reads
\begin{eqnarray}\label{Phi3}
&& |\Phi^{(L)}_5\rangle  = X_B |\Phi_4\rangle |b\rangle  |0\rangle_{\tilde B} \nonumber \\
&& =\sum_{i,j,l=1}^{N-1} \tilde a_{jl}b_i |j\rangle_R |l\rangle_C |i\rangle_b |0\rangle_{B}|0\rangle_{\tilde B}    + |g_1\rangle_{RCb}|1\rangle_{B} |0\rangle_{\tilde B} .
\end{eqnarray}
Now we refer to the Appendix,  Sec.\ref{Section:prod}, where the  matrix multiplication algorithm is described. Applying the result of this Appendix we identify $K=N$, $M=1$ and   replace the subsystems  $R_1$, $C_1$, $R_2$ with $R$, $C$, $b$, while the subsystem $C_2$ is not used because the second matrix in our multiplication is just a column. Thus,
 $W^{(1)}_{C_1R_2}$ in (\ref{W1m})  becomes  $W^{(1)}_{Cb}$,  $W^{(2)}_{C_1}$ in (\ref{W2m}) becomes $W^{(2)}_{C}$,
while the operator  $W^{(3)}_{C_1 R_2 \tilde B}$ in (\ref{W3m}) must be modified including the  projector acting on $B$, i.e. the new operator reads:
\begin{eqnarray}\label{W3mm}
P_{CBb} &=& |0\rangle_{C}|0\rangle_B |0\rangle_{b}
 \;{_{C}}\langle 0| _B\langle 0|{_b}\langle 0|, \nonumber \\
W^{(3)}_{C  Bb\tilde B} &=& P_{C Bb } \otimes \sigma^{(x)}_{\tilde B} + (I_{C Bb}- P_{CBb})\otimes I_{\tilde B}.
\end{eqnarray}
Applying the operator $W^{(3)}_{C  Bb\tilde B} W^{(2)}_{C}W^{(1)}_{Cb}$  to $|\Phi^{(L)}_5\rangle$ (\ref{Phi3}) we obtain
\begin{eqnarray}
&&
|\Phi^{(L)}_6\rangle = W^{(3)}_{CBb \tilde B} W^{(2)}_{C}W^{(1)}_{Cb}
|\Phi^{(L)}_5\rangle =\\\nonumber
&&
{\frac{1}{2^{ n/2}}}\left({\sum_{j_1=1}^{N-1}}\sum_{j=1}^{N-1}
\tilde a_{j_1 j}b_{j }  |j_1\rangle_{R} |0\rangle_{C} |0\rangle_{b} \right)|0\rangle_{B}   |1\rangle_{\tilde B}+\\\nonumber
&&  |g_2\rangle_{RCb} |1\rangle_{B}  |0\rangle_{\tilde B}.
\end{eqnarray}

Performing measurement over $\tilde B$ with the output $|1\rangle_{\tilde B}$ we remove garbage and obtain
\begin{eqnarray}\label{Phi7L}
&&
|\Phi^{(L)}_{7}\rangle =
|\Psi^{(L)}_{out}\rangle\, |0\rangle_{C} |0\rangle_{b}  |0\rangle_B ,
\\\nonumber
&&
|\Psi^{(L)}_{out}\rangle =G^{{-1}} {\sum_{j_1=1}^{N-1}}\sum_{j=1}^{N-1}
A^{-1}_{j_1 j}}b_{j }  |j_1\rangle_{R} =G^{{-1} |x\rangle_R ,
\end{eqnarray}
where   $|x\rangle_R=\sum_{j=1}^{N-1}
A^{-1}_{j_1 j}b_{j }  |j_1\rangle_{R}$ is an unnormalized vector, the normalization $G{=(\sum_{i} | x_i|^2 )^{1/2}}$,
$x_i =\sum_{j=1}^{{\color{blue}(N-1)}} A^{-1}_{i j}b_{j }$,  is known from the probability of the above measurement {which} equals $(q G \det(A))^2/2^{\tilde N +2n}$ 
due to the expression (\ref{ajl}) for $\tilde a_{jl}$. 
Thus, the probability is $\sim 1/2^{\tilde N+2n} = 2^{-O(Nn)}$. We note that, although  we use two algorithms (matrix inversion and matrix multiplication), only one measurement is performed to remove all the garbage.
The result  is stored in the register  $R$.
The depth of the whole algorithm is defined by the depth of the subroutine inverting $A$ and thus equals $O(N^2n)$. In comparison with the circuit for inverse matrix, the circuit in Fig. \ref{Fig:LinSyst}b includes $\log N$ qubits encoding the column $\boldsymbol{b}$ and thus remains essentially $O(N\log N)$.


\section{Controlled measurement}
\label{Section:CM}
In all algorithms presented above the week point is the small probability of success in the measurement of the  ancilla state. In fact, this probability is $\sim 2^{-O(N\log N)}$ in all cases,  in other words,  it exponentially decreases with an increase in the dimensionality $N$ of the matrix $\tilde A$ (\ref{AA}).  To smooth this effect we suggest to replace the usual ancilla measurement by the controlled measurement \cite{FZQW_arxive2025}, which can be presented as a special subroutine.

We call $B$ the  one-qubit ancilla whose state we have to measure and write the state of the whole system as
\begin{eqnarray}
|\chi_0\rangle = G^{-1} (|E\rangle |1\rangle_B + |F\rangle |0\rangle), \;\; \langle E|F\rangle =0,
\end{eqnarray}
where only the state  $|E\rangle$ contains the useful information, $G= \sqrt{\langle E| E\rangle + \langle F| F\rangle} $ is the normalization and the states $|E\rangle $ and $|F\rangle$ are not normalized.
First, before measurement, we double the state of  ancilla adding one more one qubit ancilla $\tilde B$ in the ground state  and applying the C-NOT $CNOT_{B\tilde B} =  |1\rangle_B\, _B\langle 1| \otimes \sigma^{(x)}_{\tilde B} +|0\rangle_B\, _B\langle 0| \otimes I_{\tilde B}$, where $\sigma^{(x)}_B$ and $I_{\tilde B}$ are, respectively, $\sigma^{(x)}$ and the identity operator applied to the ancilla $\tilde B$. We have
\begin{eqnarray}
&&
|\chi_1\rangle=CNOT_{B\tilde B} |\chi_0\rangle  |0\rangle_{\tilde B} =\\\nonumber
&& G (|E\rangle |1\rangle_B  |1\rangle_{\tilde B}+ |F\rangle |0\rangle_B |0\rangle_{\tilde B} ) .
\end{eqnarray}
Now we introduce the controlled measurement
\begin{eqnarray}\label{CM}
M_{B\tilde B} =  |1\rangle_B\, _B\langle 1| \otimes M_{\tilde B} +|0\rangle_B\, _B\langle 0| \otimes I_{\tilde B},
\end{eqnarray}
where $M_{\tilde B}$ is the measurement operator applied to $\tilde B$.  Applying the operator $M_{B\tilde B}$ to $|\chi_1\rangle$ we have
\begin{eqnarray}\label{chi2}\nonumber
|\chi_2\rangle =M_{B\tilde B} |\chi_1\rangle= |\Psi_{out}\rangle  |1\rangle_B,\; |\Psi_{out}\rangle= (\langle E| E\rangle)^{-1/2}  |E\rangle .
\end{eqnarray}
In this way we avoid necessity of performing multiple running of the algorithm  to get the needed result of the ancilla measurement.  However, we also lose the possibility to define the normalization  $(\langle E| E\rangle)^{{\color{blue}1/2}}$ which might be important information. For instance, in the algorithm for determinant calculation this normalization equals the absolute value of the matrix determinant, which is the half of useful information (another half is the phase). But in other two algorithms this normalization doesn't include the principal information. So,  getting  information about that normalization is the problem to be resolved. Another problem is the practical realization of the above controlled measurement that is also beyond of this paper. The circuit for the subroutine of controlled measurement is shown in Fig.\ref{Fig:CM}.

\begin{figure*}[!]
\centering
\subfloat[]{\includegraphics[width=5in, angle=0]{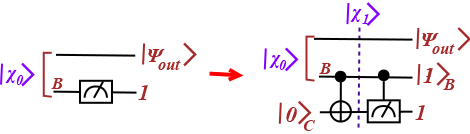}}
\caption{(Color online) Replacement of ordinary measurement of ancilla state (left circuit) with the  subroutine of controlled measurement (right circuit).}
\label{Fig:CM}
\end{figure*}

\section{Conclusions}
\label{Section:conclusions}
We propose two quantum algorithms for matrix manipulation, which are calculating the determinant and matrix inversion. Both are based on special row-wise encoding of the matrix elements in the pure states of  quantum subsystems.

The algorithm  for computing the determinant of a square matrix is discussed in details. It uses  multiqubit control-SWAP and control-Z operations. The depth of this algorithm is $O(N\log^2 N)$.  This algorithm utilized the subroutine realizing
the completely antisymmetric tensor $\varepsilon$ (operator $W^{(1)}_{SA}$) that may have relevance in other algorithms.
The outcome of this algorithm  is split into two components: the absolute value of the determinant of the matrix $A$, denoted as $|\det(A)|$, and the argument of the determinant of the matrix $A$, denoted as $\arg(\det(A))$. The first part is obtained from the measurement of an ancilla, {(although the probability of the needed outcome in such measurement decreases exponentially with $N$)}, while the argument of the determinant remains embedded in the probability amplitude of the output state $|\Psi^{(D)}_{out}\rangle$.
The size of the subsystem $S$ encoding  the matrix $A$ is $N\log N$, the size of ancillae $A_k$, $k=0,\dots,N-2$, is $\tilde N=O(N\log N)$.  Both these sizes serve to estimate the {space} of the whole algorithm which  is $O(N\log N)$ {qubits}.

It is remarkable that the $(N-1)\times (N-1)$-matrix inversion algorithm is essentially the algorithm for calculating the determinant of the
$N\times N$ matrix obtained by replacing the first  row by the row of $1/\sqrt{N}$ and the first column by the column of $q$'s (up to the diagonal element).
 This special encoding provides the superposition of all  products of matrix elements which are necessary for calculating minors included into the definition of the inverse matrix. The algorithm distributes these products with proper signs among the appropriate inverse-matrix elements which are encoded into the quantum state of certain subsystem thus forming the output state.
 The depth of the circuit is $O(N^2 \log N)$.

 As an application of inversion we develop the algorithm for solving systems of linear algebraic equations. This algorithm  consists of two blocks: matrix inversion and matrix multiplication. At that, the matrix inversion is most resource-consuming  block which finally determines the depth of the algorithm (which is $O(N^2\log N)$), and probability of required  ancilla state after measurement, $\sim 2^{-O(N\log N)}$.
 We have to note that the depths of the algorithms for calculating the determinant and matrix inverse were obtained under assumption that the $U$-, $V$-operators  in (\ref{W}) can be applied in parallel. Otherwise we have to change the depths $O(N\log^2N)$ and $O(N^2\log\,N)$ to $O(N^2\log^2N)$ in both cases.

We emphasize that the matrix-multiplication algorithm is modified in comparison with that proposed in \cite{ZQKW_arxiv2023}. The detailed description of the modified version is presented in Appendix, Sec.\ref{Section:prod}.

One notable aspect is the use of ancillae to gather all the garbage generated during the calculation, which can then be discarded with a  measurement of a one-qubit ancilla.
The weak point of  algorithms is the probability of obtaining the required state of ancilla after measuring. This probability is $\sim 2^{-O(N\log N)}$ in all three algorithms.

Combining the algorithm of matrix inversion with the algorithm of matrix multiplication we demonstrate the usage of two types of matrix encoding into the state of a quantum system. The first one is the row-wise encoding used in the input of { both  algorithm for caculating the determinant of $N\times N$ matrix and matrix inversion algorithm}  (for simplicity, we assume that $N=2^n$). It requires $N$ subsystems of $\log N$ qubits each, and the state of $i$th subsystem encodes the elements of the $i$th row  of the matrix.  The result of matrix inversion is stored in the two subsystems of $\log N$ qubits each. These  subsystems enumerate  rows and columns respectively. This form of encoding of the inverse matrix allows to perform its multiplication with the column $\boldsymbol{b}$ properly encoded in the state of $\log N$-qubit subsystem to obtain the solution $\boldsymbol{x}$ of the linear system. This solution is a column whose elements are encoded into the state of a $\log N$-qubit subsystem $R$.\\


{\bf Acknowledgements.} This project is supported by the National Natural Science Foundation of China (Grants No. 12031004, No. 12271474, and No. 61877054), and the Fundamental Research Foundation for the Central Universities (Project No. K20210337) and Zhejiang University Global Partnership Fund, 188170+194452119/003. The work was partially funded by a state task of Russian Fundamental Investigations (State Registration No. 124013000760-0).

\section{Appendix}\label{Section:appendix}
\subsection{Matrix multiplication}
\label{Section:prod}

We present an algorithm for multiplying  an $N\times K$ matrix $A^{(1)}$ by a $K\times M$  matrix $A^{(2)}$  with the elements $A^{(i)}=\{a^{(i)}_{jk}\}$, $i=1,2$,
{ assuming $N = 2^n$, $K = 2^k$, $M = 2^m$ with positive
integers $n$, $k$, $m$.} The algorithm below is an alternative to that proposed in \cite{ZQKW_arxiv2023}. Both the depth and space of the algorithm are less than those parameters in \cite{ZQKW_arxiv2023}, although the estimations for them remains the same: the depth is $O(\log K)$ and space is $O(\log K+ \log N + \log M)$ qubits.

\begin{center}
\begin{figure*}[!]
\subfloat[]{\includegraphics[width=5in, angle=0]{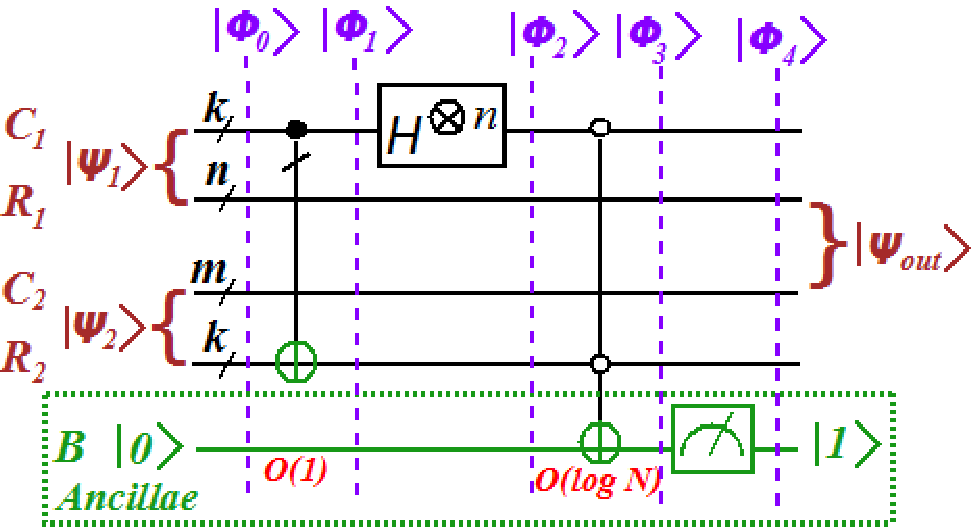}} \hspace{0.3cm}

\subfloat[]{\includegraphics[width=2.7in, angle=0]{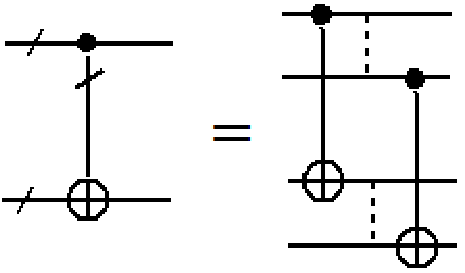}}
\caption{(a) Circuit for the matrix multiplication algorithm. We omit superscript $(m)$ in $|\Phi_k\rangle$ $(k=0,\dots,4)$ and $|\Psi_{out}\rangle$ for brevity. (b) Notation for multiqubit CNOT.}
\label{Fig:Ap1}
\end{figure*}
\end{center}

We first introduce one  register of {$n$} qubits, two registers of   {$k$} qubits and one register of {$m$} qubits which enumerate rows and columns of both matrices.
The pure states encoding the elements of matrices are
\begin{eqnarray}
|\Psi_i\rangle &=& \sum_{jl}  a^{(i)}_{jl} |j\rangle_{R_i} |l\rangle_{C_i},\;\;\;i=1,2,
\end{eqnarray}
with normalization conditions
\begin{eqnarray}\label{norm2}
\sum_{jl} |a^{(i)}_{jl}|^2=1, \;\;\;i=1,2.
\end{eqnarray}
The initial state of the whole system reads
\begin{eqnarray}
&&|\Phi_0^{(m)}\rangle =  |\Psi_1\rangle \otimes |\Psi_2\rangle=\\\nonumber
&&\sum_{j_1=0}^{N-1}\sum_{l_1,j_2=0}^{K-1}\sum_{l_2=0}^{M-1}
a^{(1)}_{j_1l_1}a^{(2)}_{j_2l_2}  |j_1\rangle_{R_1} |l_1\rangle_{C_1} |j_2\rangle_{R_2} |l_2\rangle_{C_2}.
\end{eqnarray}
From the superposition state $|\Phi_0\rangle$, we have to select terms with $l_1=j_2$.
For that purpose we apply the CNOTs $W_j$ to the $j$th qubits of $C_1$ and $R_2$,
\begin{eqnarray}
W_{j} &=& |1\rangle_{C_1,j}\; _{C_1,j}\langle 1| \otimes \sigma^{(x)}_{R_2,j} +
 |0\rangle_{C_1,j}\; _{C_1,j}\langle 0| \otimes I_{R_2,j}.
\end{eqnarray}
where $I_{R_2,j}$  is the identity operators acting on the $j$th qubit of the subsystem $R_2$.
{We have chosen the qubits of   $C_1$ as control ones. However,  if $C_1$ and $R_2$ are exchanged here and below in the description of the algorithm, it remains correct. }
 The operator $W^{(1)}_{C_1R_2}$,
\begin{eqnarray}\label{W1m}
W^{(1)}_{C_1R_2}=\prod_{j=1}^k W_j,
\end{eqnarray}
can be implemented by a depth-1 circuit, since all $W_j$ act on different qubits. Applying $W^{(1)}_{C_{1}R_2}$   to $|\Phi_0^{(m)}\rangle$ we obtain:
\begin{eqnarray}
&&
|\Phi_1^{(m)}\rangle =W^{(1)}_{C_1R_2} |\Phi_0^{(m)}\rangle
=\\\nonumber
&&{\sum_{j_1=0}^{N-1}}\sum_{j=0}^{K-1}\sum_{l_2=0}^{M-1}
a^{(1)}_{j_1 j}a^{(2)}_{j l_2}  |j_1\rangle_{R_1} |j\rangle_{C_1} |0\rangle_{R_2} |l_2\rangle_{C_2}+\\\nonumber
&&|g_1^{(m)}\rangle_{R_1C_1R_2C_2} ,
\end{eqnarray}
thus selecting terms  with  $|0\rangle_{R_2}$ from the superposition state $|\Phi_1^{(m)}\rangle$
{(the other terms are by definition garbage; their sum is denoted by
$ |g_1\rangle$ above).}
Now we apply the Hadamard transformations
\begin{eqnarray}
\label{W2m}
W^{(2)}_{C_1}=H^{\otimes  k}
\end{eqnarray}
to $C_1$  and select the terms with $|0\rangle_{C_1} |0\rangle_{R_2}$:
\begin{eqnarray}\nonumber
&&
|\Phi_2^{(m)}\rangle =W^{(2)}_{C_1}  |\Phi_1^{(m)}\rangle
 =\\\nonumber
&&
\frac{1}{2^{ k/2}}{\sum_{j_1=0}^{N-1}}\sum_{j=0}^{K-1}\sum_{l_2=0}^{M-1}
a^{(1)}_{j_1 j}a^{(2)}_{j l_2}  |j_1\rangle_{R_1} |0\rangle_{C_1} |0\rangle_{R_2} |l_2\rangle_{C_2} \\\nonumber
&& +
|g_2^{(m)}\rangle_{R_1C_1R_2C_2}.
\end{eqnarray}
Here the first term contains the desired matrix product.
Next, to label the new garbage, we prepare a one-qubit ancilla $\tilde B$ in the ground state $|0\rangle_{\tilde B}$, introduce the
projector
$
P_{C_1R_2} = |0\rangle_{C_1}|0\rangle_{R_2}
 \;{_{C_1}}\langle 0| {_{R_2}}\langle 0|
$
and the  controlled operator $W^{(3)}_{C_1 R_2 \tilde B}$,
\begin{eqnarray}
\label{W3m}
W^{(3)}_{C_1 R_2 \tilde B} = P_{C_1R_2 } \otimes \sigma^{(x)}_{\tilde B} + (I_{C_1R_2}- P_{C_1R_2})\otimes I_{\tilde B},
\end{eqnarray}
 of the depth $O(k)$ with $({2}k)$-qubit control register.
Applying this operator  to $|\Phi_2^{(m)}\rangle |0\rangle_{\tilde B}$ we obtain
\begin{eqnarray}
&&
|\Phi_3^{(m)}\rangle = W^{(3)}_{C_1 R_2 \tilde B}|\Phi_2^{(m)}\rangle |0\rangle_{\tilde B} =\\\nonumber
&&
{\frac{1}{2^{ k/2}}}\left({\sum_{j_1=0}^{N-1}}\sum_{j=0}^{K-1}\sum_{l_2=0}^{M-1}
a^{(1)}_{j_1 j}a^{(2)}_{j l_2}  |j_1\rangle_{R_1} |0\rangle_{C_1} |0\rangle_{R_2} |l_2\rangle_{C_2}\right) |1\rangle_{\tilde B}+\\\nonumber
&&  |g_2\rangle_{R_1C_1R_2C_2} |0\rangle_{\tilde B}.
\end{eqnarray}
Performing measurement over $\tilde B$ with the output $|1\rangle_{\tilde B}$, we remove the garbage and obtain
\begin{eqnarray}\nonumber
&&
|\Phi_4^{(m)}\rangle =
|\Psi_{out}^{(m)}\rangle\, |0\rangle_{C_1} |0\rangle_{R_2}  ,
\\\nonumber
&&
|\Psi_{out}^{(m)}\rangle =G^{{-1}} {\sum_{j_1=0}^{N-1}}\sum_{j=0}^{K-1}\sum_{l_2=0}^{M-1}
a^{(1)}_{j_1 j}a^{(2)}_{j l_2}  |j_1\rangle_{R_1} |l_2\rangle_{C_2},
\end{eqnarray}
where the normalization factor  $G{=(\sum_{j_1,j, l_2} | a^{(1)}_{j_1 j}a^{(2)}_{j l_2}|^2 )^{1/2}}$ is known from the probability of the above measurement {which} equals $G^2/2^{k}\sim O(2^{-k})$.
The result of the multiplication is stored in the state of the subsystem $R_1\cup C_2$.
From the above analysis we conclude that the depth  of the  whole  protocol is defined by the operator $W^{(3)}_{C_1R_2\tilde B}$ and equals  $O(k)=O(\log(K))$.
The circuit  is given in Fig. \ref{Fig:Ap1}, its space is $2 k + n + m +1$.

\subsection{Examples: determinant of $4\times 4$ matrix,  inverse of  $3\times 3$ matrix and solution of system of three linear equations}
In this section we illustrate applicability of algorithms developed in Sec.\ref{Section:det} and Sec.\ref{Section:inv} to  calculate the determinant and inverse of an arbitrary $4\times 4$ matrix assuming its invertibility.

\subsubsection{Determinant of  $4\times 4$ matrix}
\label{Section:Adet}
The generic $4\times 4$ matrix
\begin{eqnarray}\label{A_A}
\tilde A=\left(
\begin{array}{cccc}
a_{00}&a_{01}&a_{02}& a_{03}\cr
a_{10}&a_{11}&a_{12}& a_{13}\cr
a_{20}&a_{21}&a_{22}& a_{23}\cr
a_{30}&a_{31}&a_{32}& a_{33}
\end{array}
\right),
\end{eqnarray}
can be encoded into the states $|\Psi_j\rangle$,  of four two-qubit  subsystems $S_j$, $j=0,1,2,3$:
\begin{eqnarray}\label{APsi}\label{Aaa}
|\Psi_j\rangle= \sum_{k=0}^{3}a_{jk} |k\rangle_{S_0} , \;\; j=0,1,2,3,
\end{eqnarray}
with 4 obvious constraints for the  matrix elements $a_{ij}$:
\begin{eqnarray}\label{APsiNorm}
\sum_{k=0}^{3}|a_{jk}|^2=1,\;\;j=0,1,2,3.
\end{eqnarray}
The pure state of the whole system reads
\begin{eqnarray}\label{Akk}\label{APhi0}
|\Phi_0\rangle &=& |\Psi_0\rangle |\Psi_1\rangle  |\Psi_2\rangle|\Psi_{3}\rangle \nonumber\\
&=&
 \sum_{k_0,k_1, k_2, k_{3}=0}^{3} a_{0 k_0}  a_{1 k_1}   a_{2 k_2}  a_{3 k_3}  \nonumber \\
&\times &  |k_0\rangle_{S_{0}}  |k_1\rangle_{S_{1}} |k_2\rangle_{S_{2}}  |k_3\rangle_{S_{3}}   ,
\end{eqnarray}
which includes $N^{(S)}=8$ qubits.
In computing the determinant, we need only those terms in which all $k_j$ are different. Let us select those terms:
\begin{eqnarray}\label{Akk2}\label{APhi01}
|\Phi_0\rangle &=&
 \sum_{{k_0,k_1,k_2,k_3 =0}\atop{k_0\neq k_1\neq k_2\neq k_3}}^{3}  a_{0k_0} a_{1 k_1}   a_{2 k_2}  a_{3 k_3}\nonumber \\
&\times &
|k_0\rangle_{S_{0}}  |k_1\rangle_{S_{1}} |k_2\rangle_{S_{2}}  |k_3\rangle_{S_{3}}  + |g_1\rangle_S,
\end{eqnarray}
where $|g_1\rangle_S$  forms the garbage.
Introducing  ancillae  $A_k$, $k=0,1,2$ with $\tilde n_0= \tilde n_1=2$, $\tilde n_2=1$ nodes, we construct operators $W_k$ (\ref{W}):
\begin{eqnarray}\label{AW}
W_{k} = \prod_{j=k+1}^{3} U_{S_k S_{j}} V_{S_jA_k},\;\;k=0,1,2,
\end{eqnarray}
with  $U_{S_k S_{j}}$ and $V_{S_jA_k}$   given in (\ref{Vn}) and  (\ref{U}) respectively. In this case, for two-qubit ancillae $A_0$ and $A_1$ we have
$\sigma^{(x)}_{01}=\sigma^{(x)}\otimes I$, $\sigma^{(x)}_{02}=I\otimes \sigma^{(x)}$, $\sigma^{(x)}_{03}=\sigma^{(x)}\otimes \sigma^{(x)}$,  $\sigma^{(x)}_{12}=\sigma^{(x)}\otimes I$, $\sigma^{(x)}_{13}=I\otimes \sigma^{(x)}$, and for the one-qubit ancilla $A_2$ we have $\sigma^{(x)}_{23}= \sigma^{(x)}$,  where $I$ means the  one-qubit identity operator.   Also $Z_0(1) =\sigma^{(z)}\otimes I$, $Z_0(2) =Z_0(3) =I\otimes \sigma^{(z)}$, $Z_1(2) =\sigma^{(z)}\otimes I$, $Z_1(3) =I\otimes \sigma^{(z)}$,  $Z_2(3) =\sigma^{(z)}$, $\tilde N=5$.
Applying operator $W^{(1)}_{SA}$  (\ref{W1}) to the state $|\Phi_0\rangle|0\rangle_A$
 and selecting the terms which carry the information about the determinant (these terms correspond to the ground state of $A$)  we obtain ($\tilde N = 5$)
\begin{eqnarray}
&&
|\Phi_1^{(D)}\rangle =W^{(1)}_{SA} |\Phi_0\rangle |0\rangle_A \\\nonumber
&&=
 \frac{\det(\tilde A)}{2^{5/2}}\;
 |0\rangle_{S_0}|1\rangle_{S_1} |2\rangle_{S_2}  |3\rangle_{S_3}
|0\rangle_{A}+ |g_2\rangle_{SA}.
\end{eqnarray}
where
\begin{eqnarray}\label{Adet}
\det(\tilde A) &=&  \sum_{k_0\neq k_1\neq k_2\neq k_3} \epsilon_{k_0k_1k_2k_3} a_{ k_00}  a_{k_11} a_{k_22}a_{k_33}.
\end{eqnarray}
Next,  applying $\sigma^{x}$-operators to the excited qubits of $S$ we obtain:
\begin{eqnarray}
&&|\Phi_2^{(D)}\rangle=
\frac{\det(\tilde A)}{2^{5/2}}\;
 |0\rangle_{S}
|0\rangle_{A}+ |g_3\rangle_{SA} .
\end{eqnarray}
Finally, we label the garbage applying the operator  $W^{(2)}_{SAB}$ (\ref{W2})
 to $|\Phi_2^{(D)}\rangle |0\rangle_B$:
\begin{eqnarray}
&&|\Phi_3^{(D)}\rangle=
 \frac{\det(\tilde A)}{2^{5/2}}\;
 |0\rangle_{S}|0\rangle_{A}
|1\rangle_{B}+
 |g_3\rangle_{SA}|0\rangle_B .
\end{eqnarray}
and   remove the garbage  measuring  the ancilla $B$ with probability $2^{-5} |\det \tilde A|^2$ of the output $|1\rangle_{B}$  to end up with the state (\ref{Phi4D}).

For the particular matrix
\begin{eqnarray}
\tilde A=\left(
\begin{array}{cccc}
0.4&0.4&0.2& 0.8\cr
0.4i&0.2i&0.4& 0.8\cr
0.4&0.4&0.8i& 0.2i\cr
0.8&0.4&0.4&0.2
\end{array}
\right),
\end{eqnarray}
satisfying  normalization (\ref{APsiNorm}),
 the probability of success is $2^{-5} |\det \tilde A|^2=1.731\;\; 10^{-4}$, so that $|\det \tilde A|= 0.0744$ and state (\ref{Phi4D}) yields $\arg (\det(\tilde A)) =1.126$.

\subsubsection{Inverse matrix}
\label{Section:Ainv}
The $4\times 4$ matrix $\tilde A$ (\ref{tA})
reads
\begin{eqnarray}\label{AtA}
\tilde A=\left(
\begin{array}{cccc}
\frac{1}{2}&\frac{1}{2}&\frac{1}{2}&\frac{1}{2}\cr
q&a_{11}&a_{12}& a_{13}\cr
q&a_{21}&a_{22}& a_{23}\cr
q&a_{31}&a_{32}& a_{33}
\end{array}
\right).
\end{eqnarray}
Thus, we invert the $3\times 3$ matrix $A$,
\begin{eqnarray}
A=\left(
\begin{array}{ccc}
a_{11}&a_{12}& a_{13}\cr
a_{21}&a_{22}& a_{23}\cr
a_{31}&a_{32}& a_{33}
\end{array}
\right).
\end{eqnarray}
The subsystems $R$ and $C$ that enumerate rows and columns respectively are two-qubit subsystems.

Formula (\ref{Phi10}) now reads
\begin{widetext}
\begin{eqnarray}\label{APhi10}
&&
|\Phi_1\rangle =
\frac{q}{2} \sum_{j=1}^{3} \Big( \sum_{{k_2,k_3=0}\atop{0\neq k_2 \neq k_{3} \neq j }}^{3}
a_{2k_2}  a_{3 k_3}
|j\rangle_{S_0} |0\rangle_{S_1} |k_2\rangle_{S_2}  |k_3\rangle_{S_3} |j\rangle_R |1\rangle_C
+\\\nonumber
&&
\sum_{{k_1,k_3=0}\atop{0\neq k_1 \neq k_{3} \neq j}}^{3}
a_{1k_1}  a_{3 k_3}
|j\rangle_{S_0} |k_1\rangle_{S_1} |0\rangle_{S_2}  |k_3\rangle_{S_3} |j\rangle_R |2\rangle_C
+
\\\nonumber
&&
\sum_{{k_1,k_2=0}\atop{0\neq k_1 \neq k_{2} \neq j}}^{3}
a_{1k_1}  a_{2 k_2}
|j\rangle_{S_0} |k_1\rangle_{S_1} |k_2\rangle_{S_2}  |0\rangle_{S_3} |j\rangle_R |3\rangle_C
\Big)
 +  |g_2\rangle_{SRC},
\end{eqnarray}
\end{widetext}
The state $|\Phi_2$ (\ref{PHI2}) reads
\begin{eqnarray}
&&
| \Phi_2\rangle =
\frac{q  \det(A)}{2^{7/2}}
\sum_{{i=1}}^{3} \sum_{{j=1}}^{3} A^{-1}_{ji} \\\nonumber
&&\times
 |0\rangle_{S_0}|1\rangle_{S_1} |2\rangle_{S_2}  |3\rangle_{S_3}|j\rangle_R |i\rangle_C
|0\rangle_{A}+ |g_3\rangle_{SRCA} .
\end{eqnarray}
The state $|\Phi_3\rangle$ (\ref{PHI3}) reads
\begin{eqnarray}\label{APHI3}
&&|\Phi_3\rangle=
 -\frac{q  \det(A)}{2^{7/2}}\sum_{{i=1}}^{3} \sum_{{j=1}}^{3} A^{-1}_{ji}
 |0\rangle_{S}  |j\rangle_R |i\rangle_C
|0\rangle_{A}\\\nonumber
&&\times+ |g_4\rangle_{SRCA} ,
\end{eqnarray}
Finally, for the state  $|\Phi_4\rangle$ (\ref{Phi2}) we have
\begin{eqnarray}\label{APhi2}
&& |\Phi_4\rangle  =\sum_{{i=1}}^{3} \sum_{{j=1}}^{3} \tilde a_{ji}
|0\rangle_S  |j\rangle_R |i\rangle_C| 0\rangle_A |1\rangle_{B}\nonumber \\
&&+|g_5\rangle_{SRCA}|0\rangle_B,
\end{eqnarray}
where
\begin{equation} \label{Aajl}
\tilde a_{ji} = -\frac{q\det(A) A^{-1}_{ji}}{2^{7/2}}.
\end{equation}
After measuring the ancilla $B$ to get the output $|1\rangle$
 with the probability $\frac{q^2  |\det(A)|^2}{2^{7 }}  G^2 $,
 $G=\sqrt{\sum_{i,j=1}^{3}|A^{-1}_{ij}|^2 }$,
 the desired result (\ref{Phi5})
where
\begin{eqnarray}
 \label{APsiout}
|\Psi_{out}\rangle &=& G^{-1}\sum_{{i=1}}^{3} \sum_{{j=1}}^{3} A^{-1}_{ji}   |j\rangle_R |i\rangle_C .
\end{eqnarray}

For a particular matrices $\tilde A$, $A$,
\begin{eqnarray}\label{A_A2}\nonumber
\tilde A=\left(
\begin{array}{cccc}
0.5&0.5&0.5&0.5\cr
0.4&0.2&0.4& 0.8\cr
0.4&0.4&0.8& 0.2\cr
0.4&0.8&0.2&0.4
\end{array}
\right),\, A=\left(
\begin{array}{ccc}
0.2&0.4& 0.8\cr
0.4&0.8& 0.2\cr
0.8&0.2&0.4
\end{array}
\right),
\end{eqnarray}
satisfying  normalization (\ref{APsiNorm}) with $q=0.4$.
Then the probability of success is
 $\frac{q^2  |\det(A)|^2}{2^{7 }}  G^2 =1.470\;\;10^{-3}$,
 $G=2.766$.
Then $|\Psi_{out}\rangle $ yields
\begin{eqnarray}\label{AinvA}
A^{-1}=\left(
\begin{array}{ccc}
-0.714&0.000&1.429\cr
0.000&1.429& -0.714\cr
1.429&-0.714&0.000
\end{array}
\right).
\end{eqnarray}

\vspace{0.5cm}

\subsubsection{Solution of system of three linear equations}
We solve the system $A \boldsymbol{x}= \boldsymbol{b}$ with $3\times 3$ matrix $A$ given in (\ref{A_A2}).
The matrix $A^{-1}$ (\ref{AinvA})  is encoded into the state of four qubit subsystem (two-qubit subsystems $R$ and $C$), i.e., we replace $A^{-1}$ with its extended version $A^{-1}_{ext}$ and write some vector $\boldsymbol{b}$ as a vector $\boldsymbol{b}_{ext}$ with four elements:
\begin{eqnarray}\label{AinvA2}\nonumber
A^{-1}_{ext}=\left(
\begin{array}{cccc}
0&0&0&0\cr
0&-0.714&0.000&1.429\cr
0&0.000&1.429& -0.714\cr
0&1.429&-0.714&0.000
\end{array}
\right),\,\boldsymbol{b}_{ext}= \left(
\begin{array}{c}
0\cr
0.6\cr
0\cr
0.8
\end{array}
\right).
\end{eqnarray}
Thus $$\boldsymbol{x}_{ext} = A^{-1}_{ext} \boldsymbol{b}_{ext} = \left({0}\atop{\boldsymbol{x}}\right).$$
Formulae of Sec.\ref{Section:LinSyst} result in the state
$|\Psi_{out}\rangle$  presented in (\ref{Phi7L}), where
$G={\color{blue}\sqrt{\sum_{i=1}^3|x_i|^2}}=1.254$, probability of success in the ancilla measurement is
 ${\color{blue}\frac{(q G |det(A)|)^2}{2^9}}=7.546\;\;10^{-5}$,
$
\boldsymbol{x}= (0.714\;\; -0.571 \;\;0.857)^T
$
and the superscript $T$ means transpose.
}


\begin{thebibliography}{99}

\bibitem{Sho1} P. W. Shor, {\it Algorithms for quantum computation: discrete logarithms and factoring}, \href{10.1109/SFCS.1994.365700}{Proceedings 35th Annual Symposium on Foundations of Computer Science}, Santa Fe, NM, USA, pp. 124-134 (1994).

\bibitem{Sho2} P. W. Shor, {\it Polynomial-Time Algorithms for Prime Factorization and Discrete Logarithms on a Quantum Computer}, \href{https://doi.org/10.1137/S0097539795293172
} {SIAM J. Comput. {\bf 26}, 5 (1997)}.


\bibitem{Deu} D. Deutsch, {\it Quantum theory, the Church-Turing principle and the universal quantum computer}, \href{https://doi.org/10.1098/rspa.1985.0070} {Proc. R. Soc. Lond. A {\bf 400}, 97 (1985)}.


\bibitem{Gro} L. Grover, {\it A fast quantum mechanical algorithm for database search}, \href{https://doi.org/10.1145/237814.237866}{STOC '96: Proceedings 28th Annual ACM Symposium on the Theory of Computation}, ACM Press, New York (1996).

\bibitem{QFT1} D. Coppersmith, {\it An approximate Fourier transform useful in quantum factoring}, \href{https://api.semanticscholar.org/CorpusID:17450629}{IBM Research Report, RC 19642 (2002)}.

\bibitem{QFT2} Y. S. Weinstein, M. A. Pravia, E. M. Fortunato, S. Lloyd, and D. G. Cory, {\it Implementation of the Quantum Fourier Transform}, \href{https://doi.org/10.1103/PhysRevLett.86.1889} {Phys. Rev. Lett. {\bf 86}, 1889 (2001)}.

\bibitem{QFT3} M. A. Nielsen and I. L. Chuang, {\it Quantum Computation and Quantum Information}, Cambridge University Press, Cambridge, England (2000).

\bibitem{Wang} H. Wang, L. Wu, Y. Liu, and F. Nori,
{\it Measurement-based quantum phase estimation algorithm for finding eigenvalues of non-unitary matrices}, \href{https://doi.org/10.1103/PhysRevA.82.062303}
{Phys. Rev. A. {\bf 82}, 062303 (2010)}.

\bibitem{HHL} A. W. Harrow, A. Hassidim, and S. Lloyd, {\it Quantum Algorithm for Linear Systems of Equations}, \href{https://doi.org/10.1103/PhysRevLett.103.150502} {Phys. Rev. Lett. {\bf 103}, 150502 (2009)}.

\bibitem{HHL1} B. D. Clader, B. C. Jacobs, and C. R. Sprouse, {\it Preconditioned Quantum Linear System Algorithm}, \href{https://doi.org/10.1103/PhysRevLett.110.250504} {Phys. Rev. Lett. {\bf 110}, 250504 (2013)}.

\bibitem{HHL2} J. Biamonte, P. Wittek, N. Pancotti, P. Rebentrost, N. Wiebe, and S. Lloyd,
{\it Quantum machine learning}, \href{https://doi.org/10.1038/nature23474} {Nature. {\bf 549}, 195 (2017)}.

\bibitem{HHL3} L. Wossnig, Z. Zhao, and A. Prakash,
{\it Quantum Linear System Algorithm for Dense Matrices}, \href{https://doi.org/10.1103/PhysRevLett.120.050502}
{Phys. Rev. Lett. {\bf 120}, 050502 (2018)}.

\bibitem{HHL4} X. D. Cai, C. Weedbrook, Z. E. Su, M. C. Chen, M. Gu, M. J. Zhu, L. Li, N. L. Liu, C. Y. Lu, and J.W. Pan, {\it Experimental Quantum Computing to Solve Systems of Linear Equations},
\href{https://doi.org/10.1103/PhysRevLett.110.230501} {Phys. Rev. Lett. {\bf 110}, 230501 (2013)}.

\bibitem{HHL5} J. W. Pan, Y. Cao, X. Yao, Z. Li, C. Ju, H. Chen, X. Peng, S. Kais, and J. Du,
{\it Experimental realization of quantum algorithm for solving linear systems of equations}, \href{https://doi.org/10.1103/PhysRevA.89.022313}
{Phys. Rev. A. {\bf 89}, 022313 (2014)}.

\bibitem{HHL6} S. Barz, I. Kassal, M. Ringbauer, Y. O. Lipp, B. Daki\'{c}, A. Aspuru-Guzik, and P. Walther, {\it A two-qubit photonic quantum processor and its application to solving systems of linear equations}, \href{https://doi.org/10.1038/srep06115} {Sci. Rep. {\bf 4}, 6115 (2014)}.

\bibitem{HHL7} Y. Zheng, C. Song, M. C. Chen, B. Xia, W. Liu, Q. Guo, L. Zhang, D. Xu, H. Deng, K. Huang, Y. Wu, Z. Yan, D. Zheng, L. Lu, J. W. Pan, H. Wang, C. Y. Lu, and X. Zhu,
{\it Solving Systems of Linear Equations with a Superconducting Quantum Processor}, \href{https://doi.org/10.1103/PhysRevLett.118.210504}
{Phys. Rev. Lett. {\bf 118}, 210504 (2017)}.

\bibitem{T}
 A. T.-Shma, {\it Inverting well conditioned matrices in quantum logspace},
in Proceedings of the Forty-fifth Annual ACM Symposium on Theory of Computing, STOC '13, pages 881-890, New York, USA (2013).

\bibitem{AT}
D. Aharonov and A. T.-Shma, {\it Adiabatic quantum state generation}, \href{https://doi.org/10.1137/060648829} {SIAM Journal on Computing
{\bf 37}(1), 47 (2007)}.

\bibitem{BHMT}
G. Brassard, P. Hoyer, M. Mosca, and A. Tapp, {\it Quantum amplitude amplification and estimation}, \href{https://doi.org/10.1090/conm/305/05215} {Contemp. Math. {\bf 305},  53-74 (2002)}.

\bibitem{CKS}
A. M. Childs, R. Kothari, and R. D. Somma, {\it Quantum algorithm for systems of linear equations
with exponentially improved dependence on precision}, \href{https://doi.org/10.1137/16M1087072}{SIAM J. Comput. {\bf 46}, 1920-1950 (2017)}.

\bibitem{GSLW}
A. Gily\'en, Y. Su, G. H. Low, and N. Wiebe, {\it Quantum singular value transformation and
beyond: exponential improvements for quantum matrix arithmetics}, \href{https://doi.org/10.1145/3313276.3316366}{STOC 2019: Proceedings of the 51st Annual ACM SIGACT Symposium on Theory of Computing, 193-204 (2019)}.

\bibitem{LC}
G. H. Low and I. L. Chuang, {\it Optimal Hamiltonian simulation by quantum signal processing},
\href{https://doi.org/10.1103/PhysRevLett.118.010501} {Phys. Rev. Lett. {\bf 118}, 010501 (2017)}.

\bibitem{LYC}
G. H. Low, T. J. Yoder, and I. L. Chuang, {\it Methodology of resonant equiangular composite
quantum gates}, \href{ https://doi.org/10.1103/PhysRevX.6.041067} {Phys. Rev. X {\bf 6}, 041067 (2016)}.

\bibitem{AL}
D. An and L. Lin, {\it Quantum linear system solver based on time-optimal adiabatic quantum
computing and quantum approximate optimization algorithm}, \href{https://doi.org/10.1145/349833} {ACM Transactions on Quantum Computing {\bf 3( 2)}, Article 5, pages 1-28 (2022)}.

\bibitem{JRS}
S. Jansen, M.-B. Ruskai, and R. Seiler, {\it Bounds for the adiabatic approximation with applications to quantum computation}, \href{https://doi.org/10.1063/1.2798382} {J. Math. Phys. {\bf 48}(10), 102111 (2007)}.

\bibitem{SSO}
Y. Suba\c{s}\i, R. D. Somma, and D. Orsucci, {\it Quantum algorithms for systems of linear equations inspired by adiabatic quantum computing}, \href{https://doi.org/10.1103/PhysRevLett.122.060504} {Phys. Rev. Lett. {\bf 122}, 060504 (2019)}.

\bibitem{AlL}
 T. Albash and D. A. Lidar, {\it Adiabatic quantum computation}, \href{https://doi.org/10.1103/RevModPhys.90.015002} {Rev. Mod. Phys. {\bf 90}, 015002
(2018)}.

\bibitem{TAWL}
Y. Tong, D. An,  N. Wiebe, and L. Lin,
{\it Fast inversion, preconditioned quantum linear system solvers, fast Green's-function
computation, and fast evaluation of matrix functions}, \href{https://doi.org/10.1103/PhysRevA.104.032422} {Phys. Rev. A {\bf 104}, 032422 (2021)}.

\bibitem{A}
S. Aaronson, {\it Read the fine print},  \href{https://doi.org/10.1038/nphys3272} {Nature Phys. {\bf 11}, 291-293 (2015)}.

\bibitem{RML}
P. Rebentrost,, M. Mohseni, and S. Lloyd, {\it Quantum Support Vector Machine for Big Data Classification}, \href{https://doi.org/10.1103/PhysRevLett.113.130503} {Phys. Rev. Lett. {\bf 113}, 130503 (2014)}.

\bibitem{WBL}
N. Wiebe, D. Braun, and S. Lloyd,
{\it Quantum Algorithm for Data Fitting}, \href{https://doi.org/10.1103/PhysRevLett.109.050505} {Phys. Rev. Lett. {\bf 109}, 050505 (2012)}.

\bibitem{SSP}
M. Schuld, I. Sinayskiy, and F. Petruccione, {\it Prediction by linear regression on a quantum computer}, \href{https://doi.org/10.1103/PhysRevA.94.022342} {Phys. Rev.  A {\bf  94}, 022342 (2016)}.

\bibitem{Wang2}
G. Wang, {\it Quantum algorithm for linear regression}, \href{https://doi.org/10.1103/PhysRevA.96.012335} {Phys. Rev. A {\bf 96}, 012335 (2017)}.

\bibitem{G}
F. L. Gall, {\it Robust Dequantization of the Quantum Singular value Transformation and
Quantum Machine Learning Algorithms}, \href{https://doi.org/10.48550/arXiv.2304.04932} {arXiv:2304.04932 [quant-ph]}.


\bibitem{B}
D. W. Berry, {\it High-order quantum algorithm for solving linear differential equations},  \href{https://doi.org/10.1088/1751-8113/47/10/105301} {J. Phys. A: Math. Theor. {\bf 47}(10) 105301 (2014)}.

\bibitem{LKKLTC}
J.-P. Liu, H.  Kolden, H. K. Krovi,  N. F. Loureiro, K.Trivisa, and A. M. Childs,
{\it Efficient quantum algorithm for dissipative nonlinear differential equations}, \href{https://doi.org/10.1073/pnas.202680511} {PNAS {\bf 118}(35): e2026805118 (2021)}.

\bibitem{MRTC}
J.M. Martyn, Z. M. Rossi, A.K. Tan, I.L. Chuang,
A Grand Unification of Quantum Algorithms, PRX Quantum {\bf 2}, 040203 (2021)

\bibitem{NJ}
I. Novikau, I. Joseph
Estimating QSVT angles for matrix inversion with large condition numbers, arXiv:2408.15453v2 [quant-ph]  (2024)


\bibitem{ZhaoL1} L. Zhao, Z. Zhao, P. Rebentrost, and J. Fitzsimons,
{\it Compiling basic linear algebra subroutines for quantum computers}, \href{https://doi.org/10.1007/s42484-021-00048-8} {Quantum Mach. Intell. {\bf 3}, 21 (2021)}.

\bibitem{DFZ_2020}
S. I. Doronin, E. B. Fel'dman, and A. I. Zenchuk,
{\it Solving systems of linear algebraic equations via unitary
transformations on quantum processor of IBM Quantum
Experience}, \href{https://doi.org/10.1007/s11128-019-2570-5}{Quantum Inf. Process {\bf 19},  68 (2020)}.


\bibitem{LWWZ}
H. Li, N. Jiang, Z. Wang, J. Wang, and R. Zhou, {\it Quantum Matrix Multiplier}, \href{https://doi.org/10.1007/s10773-021-04816-x}{Int. J. Theor. Phys. {\bf 60}, 2037-2048 (2021)}.


\bibitem{KN}
R. Kothari and A. Nayak,  {\it Quantum Algorithms for Matrix Multiplication and Product Verification}, \href{https://doi.org/10.1007/978-3-642-27848-8_303-2}{In: M.-Y. Kao (eds), Encyclopedia of Algorithms, Springer, Berlin, Heidelberg (2015)}.



 \bibitem{QZKW_arxive2022}
 W. Qi, A. I. Zenchuk, A. Kumar, J. Wu,
{\it Quantum algorithms for matrix operations and linear systems of equations}, \href{https://doi.org/10.1088/1572-9494/ad2366}{Commun. Theor. Phys. {\bf 76}, 035103 (2024)}.


\bibitem{ZQKW_arxiv2023}
A. I. Zenchuk, W. Qi, A. Kumar, and J. Wu,
{\it Matrix manipulations via unitary transformations and ancilla-state measurements},
\href{https://www.rintonpress.com/xxqic24/qic-24-1314/1099-1109.pdf} {Quantum Inf. Comp. {\bf 24(13,14)}, 1099-1109 (2024)}.

\bibitem{Ber}
S. J. Berkowitz, {\it On computing the determinant in small parallel time using a small number
of processors},  \href{https://doi.org/10.1016/0020-0190(84)90018-8} {Inf. Process. Lett. {\bf 18(3)}, 147-150 (1984)}.


\bibitem{BAEM}
E. Boix-Adser\'a, L. Eldar, S. Mehraban,
{\it Approximating the Determinant of Well-Conditioned Matrices by Shallow Circuits},
\href{https://doi.org/10.48550/arXiv.1912.03824} {arXiv:1912.03824 [cs.DS]}.

\bibitem{KShV}
A.Yu. Kitaev, A. H. Shen, M. N. Vyalyi, {\it Classical and Quantum Computation}, Graduate
Studies in Mathematics, V.47, American Mathematical Society, Providence, Rhode Island
(2002).


\bibitem{FZQW_arxive2025}
E. B. Fel’dman, A. I. Zenchuk, W. Qi,  and J. Wu, Remarks on controlled measurement and quantum algorithm for calculating Hermitian conjugate, 	arXiv:2501.16028v1.

\end{thebibliography}
\end{document}